\documentclass[nofootinbib,eqsecnum,superscriptaddress,amsmath,amssymb,aps,11pt]{revtex4-2}

\usepackage{amsfonts}
\usepackage{bbm} 

\usepackage{pict2e} 
\usepackage{graphpap} 

\usepackage{tikz}
\usetikzlibrary{quantikz}

\usepackage{footnotebackref} 

\DeclareMathOperator{\Tr}{Tr} 

\newcommand{\Path}[1]{{\mathrm{Path}_{#1}}} 

\newcommand{\secref}[1]{Sec.~\ref{#1}}

\newcommand{\figref}[1]{Fig.~\ref{#1}}
\newcommand{\appref}[1]{Appendix~\ref{#1}}

\begin{document}
\count\footins = 1000 

\title{Demonstration of Scully--Dr{\"u}hl-type quantum erasers on quantum computers}



\author{Bo-Hung Chen}
\email{kenny81778189@gmail.com}
\thanks{}
\affiliation{Department of Physics, National Taiwan University, Taipei 10617, Taiwan}
\affiliation{Graduate Institute of Electronics Engineering, National Taiwan University, Taipei 10617, Taiwan}
\affiliation{Center for Quantum Science and Engineering, National Taiwan University, Taipei 10617, Taiwan}
\affiliation{Physics Division, National Center for Theoretical Sciences, Taipei 10617, Taiwan}

\author{Dah-Wei Chiou}
\email{dwchiou@gmail.com}
\thanks{}
\affiliation{Graduate Institute of Electronics Engineering, National Taiwan University, Taipei 10617, Taiwan}
\affiliation{Center for Quantum Science and Engineering, National Taiwan University, Taipei 10617, Taiwan}
\affiliation{Physics Division, National Center for Theoretical Sciences, Taipei 10617, Taiwan}

\author{Hsiu-Chuan Hsu}
\email{hcjhsu@nccu.edu.tw}
\thanks{The authors contributed equally.}
\affiliation{Graduate Institute of Applied Physics, National Chengchi University, Taipei 11605, Taiwan}
\affiliation{Department of Computer Science, National Chengchi University, Taipei 11605, Taiwan}


\begin{abstract}
We present a novel quantum circuit that genuinely implements the Scully–Dr{\"u}hl-type delayed-choice quantum eraser, where the two recorders of the which-way information directly interact with the signal qubit and remain spatially separated. Experiments conducted on IBM Quantum and IonQ processors demonstrate that the recovery of interference patterns, to varying degrees, aligns closely with theoretical predictions, despite the presence of systematic errors. This quantum circuit-based approach, more manageable and versatile than traditional optical experiments, facilitates arbitrary adjustment of the erasure and enables a true random choice in a genuine delayed-choice manner. On the IBM Quantum platform, delay gates can be employed to further defer the random choice, thereby amplifying the retrocausal effect. Since gate operations are executed sequentially in time, the system does not have any involvement of random choice until after the signal qubit has been measured, therefore eliminating any potential philosophical loopholes regarding retrocausality that might exist in other experimental setups. Remarkably, quantum erasure is achieved with delay times up to $\sim1\,\mu\text{s}$ without noticeable decoherence, a feat challenging to replicate in optical setups.
\end{abstract}

\maketitle


\section{Introduction}

The quantum eraser is an interferometer experiment in which the which-way information of each single quanton (i.e., quantum particle such as photon) is ``marked'' in the first place but can be ``erased'' later. As the which-way information is marked, the wave property is not manifested and thus the interference pattern is not seen. However, the interference pattern can be ``recovered'' if the which-way information is erased. The erasure can be made even after the quanton has been detected by the detector. In a sense, the behavior of the quanton in the past can be \emph{retroactively} affected by a \emph{delayed-choice} measurement performed in a later time.
The original idea of the quantum eraser was proposed by Scully and Dr{\"u}hl in 1982 \cite{PhysRevA.25.2208}. The first experimental realization was performed by Kim \textit{et al.}\ in 1999 \cite{PhysRevLett.84.1} in a double-slit interference experiment. A similar double-slit experiment featuring entanglement of photon polarization was later performed by Walborn \textit{et al.}\ in 2002 \cite{PhysRevA.65.033818}.
To date, many more quantum eraser experiments under different scenarios have been proposed and many of them have been experimentally realized (see \cite{RevModPhys.88.015005} for a review).

Ever since the delayed-choice quantum eraser was proposed, its interpretations and implications have been the subject of vigorous debate that has persisted to this day \cite{englert1999quantum, mohrhoff1999objectivity, aharonov2005time, hiley2006erased, ellerman2015delayed, fankhauser2017taming, kastner2019delayed, qureshi2020}.
In particular, drawing an analogy to the Einstein--Podolsky--Rosen--Bohm (EPR--Bohm) experiment \cite{PhysRev.108.1070,RevModPhys.81.1727}, Kastner argued that the quantum eraser neither erases nor delays any information, thus displaying no mystery at all beyond the standard EPR correlation \cite{kastner2019delayed}.
Subsequently, recasting the quantum eraser in a Mach--Zehnder interferometer, Qureshi further elaborated on the analogy to the EPR--Bohm experiment and contended that the quantum erasure exhibits no retrocausal effect whatsoever \cite{qureshi2020}.
Furthermore, by considering the Mach--Zehnder interferometer generalized with a nonsymmetric beam splitter, it was explicitly shown that the quantum eraser shares exactly the same formal (i.e., mathematical) structure with the
EPR--Bohm experiment and thus can be understood in terms of the EPR correlation \cite{chiou2023delayed}. Nevertheless, the quantum eraser still poses a conceptual issue beyond the standard EPR paradox \cite{chiou2023delayed}, as opposed to what is claimed otherwise, e.g., in \cite{kastner2019delayed, qureshi2020}.

According to the detailed analysis of \cite{chiou2023delayed}, the quantum eraser experiments can be classified into two conceptually rather different categories: the \emph{entanglement quantum eraser} and the \emph{Scully--Dr{\"u}hl-type quantum eraser}.

The entanglement quantum eraser is based on the entanglement of some internal states between a pair of quantons (referred to as the \emph{signal} and \emph{idler} quantons). The experiment performed by Walborn \textit{et al.}\ \cite{PhysRevA.65.033818} is a typical example, which involves the entanglement of polarization between a pair of photons. In the entanglement quantum eraser, the which-way information of the signal quanton is ``marked'' in terms of some \emph{internal state} of the idler quanton (e.g., polarization of the idler photon), which can be either read out or erased by different delayed-choice settings of measurement. However, because the which-way information of the signal quanton is inferred from the internal state of the idler quanton, of which the measurement does not invoke any direct contact with the signal quanton, the inference is \emph{counterfactual} in nature and thus it can always be argued that the which-way information of the signal quanton is not marked in the first place and not erased in a later time, if counterfactual reasoning is all dismissed.

On the other hand, in the Scully--Dr{\"u}hl-type quantum eraser as originally proposed by Scully and Dr{\"u}hl \cite{PhysRevA.25.2208} and performed by Kim \textit{et al.}\ \cite{PhysRevLett.84.1}, the which-way information of the signal quanton is ``marked'' in terms of the states of two objects that are in \emph{direct contact} with the traveling paths of the signal quanton. Because the two objects serving as the ``recorders'' are in direct contact with the paths, the which-way information inferred from measuring the states of the two recorders becomes \emph{factual} if the measurement yields a conclusive outcome.
Therefore, the assertion that the which-way information can be influenced by the delayed-choice measurement, even retroactively, is not just a consequence of counterfactual reasoning but bears some factual significance.
In this sense, the Scully--Dr{\"u}hl-type quantum eraser presents a ``mystery'' deeper than the entanglement quantum eraser.
Furthermore, it is even more remarkable that the two recorders are \emph{spatially separated} in the first place, yet the record can still be erased.

The distinction between these two types of quantum erasers is both fundamental and significant, irrespective of philosophical interpretation. Experimentally examining both is vital for probing the foundations of quantum mechanics, as a deeper, more fundamental theory beyond standard quantum mechanics could, in principle, confirm the predictions of quantum mechanics for one type while deviating from them for the other.
Aside from the original concept proposed by Scully and Dr{\"u}hl \cite{PhysRevA.25.2208} and its experimental realization by Kim \textit{et al.}\ \cite{PhysRevLett.84.1}, most theoretical models and experimental implementations of the quantum eraser either fall into the category of the entanglement quantum eraser or differ significantly from the Scully--Dr{\"u}hl type (see \cite{RevModPhys.88.015005}). Conducting more experiments of the genuine Scully--Dr{\"u}hl type is of great importance to further test the foundations of quantum mechanics with regard to retrocausality.

Meanwhile, by considering the interference between two orthogonal qubit states, the delayed-choice experiments can be redesigned into quantum circuits, which offer a higher level of abstraction as the information flow becomes more transparent \cite{PhysRevLett.107.230406}.
Today, various cloud services of quantum computing, such as IBM Quantum \cite{IBMQ} and Amazon Braket \cite{AWS}, provide accessible and easily manageable facilities for performing quantum experiments. In the last few years, the IBM Quantum platform has been used to perform intriguing interference experiments \cite{PhysRevA.102.032605, PhysRevA.103.022409, PhysRevA.103.022212, PhysRevA.104.032223, tran2022experimenting}. Particularly, the recent work of \cite{chiou2024complementarity} implements an entanglement quantum eraser on the quantum computers of IBM Quantum with the extension that the degree of entanglement is adjustable.

In this paper, in the same spirit of \cite{chiou2024complementarity}, we propose a quantum circuit that genuinely realizes the Scully--Dr{\"u}hl-type quantum eraser and perform the experiments both in the quantum processor of superconducting transmons on the IBM Quantum platform and in the trapped ion quantum processor IonQ \cite{IonQ} on the Amazon Braket platform. The hardware architectures of the quantum processors of IBM Quantum and the IonQ processor ensure that the two qubits recording the which-way information remain spatially separated by distances on the order of $\sim10^2\mu\text{m}$ \cite{chow2015characterizing} and $\sim1\,\mu\text{m}$, respectively \cite{IonQ, doi:10.1126/science.1231298}. These experiments provide more experimental realizations of the genuine Scully--Dr{\"u}hl-type quantum eraser using superconducting transmons and trapped ions.

The quantum circuit experiment is not only easier to implement than optical experiments, but also offers the advantage that the degree of erasure is readily adjustable, making it easy to obtain full or partial recovery of the interference pattern to any desired degree, a feat that is rather difficult to achieve in optical experiments.
Moreover, optical experiments are often plagued by various unwanted sources of decoherence. For example, in the double-slit experiment of the Scully–Dr{\"u}hl-type quantum eraser performed by Kim \textit{et al.}\ \cite{PhysRevLett.84.1}, the finite width of the slits introduces additional degrees of decoherence, significantly reducing the contrast of the recovered interference pattern compared to the optimal case. In contrast, the experiments conducted on both IBM Quantum and IonQ platforms allow for nearly complete recovery of the interference pattern, with well-defined nodes and antinodes, thanks to the high fidelity of the quantum circuit hardware achieved by state-of-the-art technology.

In a quantum circuit, we can automate the random choice of deciding the degree of erasure by utilizing the quantum randomness of ancilla qubits. Unlike pseudorandom sources, qubits provide true randomness, ensuring that the decision is genuinely random and not predetermined in any way. As the choice made remains unknown until the ancilla state is read out, this random choice can be considered to be made in a delayed-choice manner.
Furthermore, on IBM Quantum processors, the measurement of the signal qubit can be performed midway, ensuring that the random choice is genuinely invoked in a delayed-choice manner (i.e., after the signal qubit has been measured).
Additionally, \emph{delay gates} can be employed to further postpone the random choice, extending the deferral and potentially amplifying the retrocausal effect.

In the IBM Quantum framework, since gate operations are executed sequentially in time, there is no involvement of random choice at all before the signal qubit is measured. This approach contrasts markedly with the experimental setup in \cite{PhysRevLett.84.1}, where the random choice is made by two beam splitters that are continuously present.
Philosophically, one could argue that the retrocausal effect demonstrated in \cite{PhysRevLett.84.1} arises simply because the random choice devices (i.e., the two beam splitters) were prearranged in advance. In our experiments on the IBM Quantum platform, such a philosophical loophole is absent, thereby enhancing the retrocausal nature of the delayed-choice quantum eraser.\footnote{\label{foot:loophole}Questions such as whether the retrocausal effect is enhanced are philosophical in nature and not the main focus of this work. Nevertheless, the distinction between scenarios where the philosophical loophole persists and those where it is eliminated is a fundamental difference in experimental implementation. For testing the foundations of quantum mechanics, it is crucial to experimentally push the boundaries to tighten such loopholes.}

This paper is organized as follows.
In \secref{sec:quantum eraser}, we briefly review the ideas and concepts of the delayed-choice quantum eraser, emphasizing the differences between the entanglement quantum eraser and the Scully--Dr{\"u}hl-type quantum eraser.
In \secref{sec:quantum circuit}, we propose an implementation of the Scully--Dr{\"u}hl-type quantum eraser in quantum circuits.
In \secref{sec:IBM} and \secref{sec:IonQ}, we present the experimental results conducted on IBM Quantum and IonQ, respectively.
Finally, we summarize and discuss the results and their implications in \secref{sec:summary}.
More experimental data in different settings are presented in Appendix~\ref{app:no choice} and Appendix~\ref{app:4 options}. Some technical details are provided in Appendix~\ref{app:technical details}.

\section{Delayed-choice quantum eraser: ideas and concepts}\label{sec:quantum eraser}
In this section, we briefly overview the ideas and concepts of the delayed-choice quantum eraser. We first present the entanglement quantum eraser and then the Scully--Dr{\"u}hl-type quantum eraser, following the same line as in \cite{chiou2023delayed}.\footnote{The work of \cite{chiou2023delayed} considers the quantum erasers with extensions that render the quantum erasers mathematically equivalent to the EPR--Bohm experiment. Here, for our purpose, we do not include these extensions. Also note that the notations $\theta$, $\phi$, etc.\ used here are different from those in \cite{chiou2023delayed}.} The quantum eraser experiment is formulated in terms of a Mach--Zehnder interferometer, which is conceptually more concise than a double-slit experiment and draws a direct analogy implementable in a quantum circuit. More details can be found in \cite{chiou2023delayed}; also see \cite{RevModPhys.88.015005} for a comprehensive review of quantum erasers in general.

\subsection{Entanglement quantum eraser}
\begin{figure}
\centering
    \includegraphics[width=0.85\textwidth]{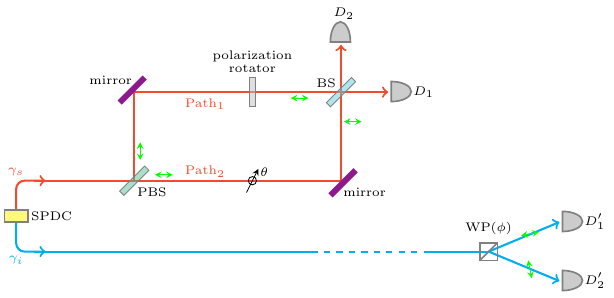}
\caption{The schematic diagram of an entanglement quantum eraser. A pair of entangled photons $\gamma_s$ and $\gamma_i$ with orthogonal polarizations are created by SPDC. The signal photon $\gamma_s$ is directed into a Mach--Zehnder interferometer with the detectors $D_1$ and $D_2$. The idler photon $\gamma_i$ is directed into the delayed-choice measuring device with the detectors $D'_1$ and $D'_2$.}
\label{fig:interferometer}
\end{figure}

The optical experiment of an entanglement quantum eraser is illustrated in \figref{fig:interferometer}.\footnote{The idea of Fig.~1 in \cite{qureshi2021delayed} is adopted here for the interferometer. The concept of \figref{fig:interferometer} is the same as that of Fig.~1 in \cite{chiou2024complementarity}, except that a different method is used to recombine the two paths.} Spontaneous parametric down-conversion (SPDC) in a nonlinear optical crystal is used to prepare a pair of photons (referred to as the \emph{signal} photon $\gamma_s$ and the \emph{idler} photon $\gamma_i$) that are entangled with orthogonal polarizations. The signal photon $\gamma_s$ is directed into a Mach--Zehnder interferometer with the detectors $D_1$ and $D_2$, while the idler photon $\gamma_i$ is directed into the ``delayed-choice'' measuring device with the detectors $D'_1$ and $D'_2$.
As $\gamma_s$ enters the interferometer, it is split by the polarizing beam splitter $\mathrm{PBS}$ into two the paths, $\Path{1}$ and $\Path{2}$, with horizontal ($\leftrightarrow$) and vertical ($\updownarrow$) polarizations, respectively.
Along $\Path{2}$, an adjustable phase shift $\theta$ is introduced (e.g, by inserting a phase-shift plate).
Along $\Path{1}$, a polarization rotator that rotates $\updownarrow$ into $\leftrightarrow$ is introduced in order to make the two paths interfere with each other.
The two paths are finally recombined by the beam splitter $\mathrm{BS}$ before $\gamma_s$ clicks either of the two detectors, $D_1$ and $D_2$.

Meanwhile, the idler photon $\gamma_i$ is directed into a Wollaston prism that splits two mutually orthogonal polarizations into the detectors $D'_1$ and $D'_2$ separately. The orientation of the Wollaston prism can be adjusted. At the orientation angle described by $\phi$, the linear polarization at the angle $\phi$ from the horizontal direction enters $D'_1$ while the orthogonal linear polarization at the angle $\pi/2+\phi'$ enters $D'_2$.
The Wollaston prism can be located more distant away from the SPDC source than $D_1$ and $D_2$, so that the value of $\phi$ can be adjusted in the ``delayed-choice'' manner \emph{after} the signal photon $\gamma_s$ has already registered a signal in $D_1$ or $D_2$.

If we repeat the experiments numerous times, we obtain the detection probabilities of $D_1$ and $D_2$ from the accumulated counts of individual signals. Since $\gamma_s$ and $\gamma_i$ are maximally entangled in polarization, $\gamma_s$ is completely unpolarized and can be said to travel either $\Path{1}$ or $\Path{2}$ with equal probability. As a result, the detection probabilities of $D_1$ and $D_2$ are 50\% for each, displaying no interference pattern in response to the adjustable phase shift $\theta$.

Meanwhile, we can group the accumulated events of $\gamma_s$ into to two subensembles according to whether the corresponding $\gamma_i$ clicks $D'_1$ or $D'_2$. Each individual event of $\gamma_s$ in the subensembles associated with $D'_1$ and $D'_2$ is linearly polarized at the angle $\pi/2+\phi$ and $\phi$ respectively, as a consequence of the polarization entanglement between $\gamma_s$ and $\gamma_i$.

If we set $\phi=0$, the polarization of $\gamma_s$ in the $D'_1$-subensemble and $D'_2$-subensemble is $\updownarrow$ and $\leftrightarrow$ respectively, which correspond to $\Path{1}$ and $\Path{2}$ separately. The which-way information of $\gamma_s$ is said to be ``marked'' by the polarization state of $\gamma_i$. As the which-way information of whether each individual $\gamma_s$ travels $\Path{1}$ or $\Path{2}$ is marked, within the confines of each subensemble, the detection probabilities remains independent of $\theta$.
If we set $\phi=\pi/2$, the situation is the same except that now $D'_1$ corresponds to $\Path{1}$ and $D'_2$ corresponds to $\Path{2}$.

However, if we set $\phi=\pi/4$, the outcomes of $D'_1$ and $D'_2$ correspond to $(\ket{\leftrightarrow}+\ket{\updownarrow})/\sqrt{2}$ and $(\ket{\leftrightarrow}-\ket{\updownarrow})/\sqrt{2}$ respectively for the polarization of $\gamma_i$, which in turn correspond to $(\ket{\leftrightarrow}-\ket{\updownarrow})/\sqrt{2}$ and $(\ket{\leftrightarrow}+\ket{\updownarrow})/\sqrt{2}$ respectively for the polarization of $\gamma_s$. Consequently, the which-way information of each individual $\gamma_i$ is completely ``erased'' by the associated $D'_1$ or $D'_2$ outcome, and each $\gamma_s$ is said to travel \emph{both} paths simultaneously. Accordingly, within each subensemble associated with $D'_1$ or $D'_2$, the two-path interference pattern is fully ``recovered'' in the sense that the detection probabilities of $D_1$ and $D_2$ appear modulated in response to $\theta$.

Furthermore, If we adjust $\phi$ to some angle other than $0$, $\pi/2$, and $\pi/4$, the which-way information of each $\gamma_s$ is partially erased to a certain degree. Correspondingly, within each subensemble associated with $D'_1$ or $D'_2$, the interference pattern is partially recovered as the detection probabilities of $D_1$ and $D_2$ appear as partially modulated in response to $\theta$.

As the value of $\phi$ can be adjusted in the delayed-choice manner, whether each $\gamma_s$ travels along $\Path{1}$, $\Path{2}$, or both apparently can be affected, even \emph{retroactively}, by the measurement of $\gamma_i$ performed in a later time. It is this effect of \emph{retrocausality} that has aroused fierce controversy \cite{englert1999quantum, mohrhoff1999objectivity, aharonov2005time, hiley2006erased, ellerman2015delayed, fankhauser2017taming, kastner2019delayed, qureshi2020}.

The which-way information of $\gamma_s$ is inferred from the polarization state of $\gamma_i$ through entanglement. Because the measurement of the polarization of $\gamma_i$ does not invoke any direct contact with $\gamma_s$, the inference about the which-way information of $\gamma_s$ is \emph{counterfactual} in nature.
If counterfactual reasoning is completely dismissed, one can insists that the which-way information of $\gamma_s$ is not marked in the first place and not erased later. In this viewpoint, the entanglement quantum eraser does not present any additional mystery beyond the standard EPR paradox.

\subsection{Scully--Dr{\"u}hl-type quantum eraser}\label{sec:SD quantum eraser}
\begin{figure}
\centering
    \includegraphics[width=0.85\textwidth]{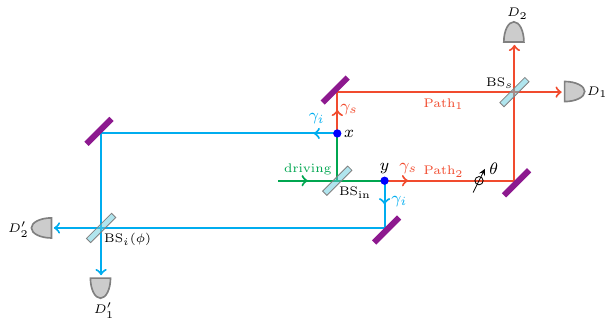}
\caption{The schematic diagram of a Scully--Dr{\"u}hl-type quantum eraser. Driving pulses enter the beam splitter $\mathrm{BS_{in}}$ and trigger either of the two objects $x$ and $y$ to produce a pair of photons $\gamma_s$ and $\gamma_i$. The photon $\gamma_s$ is directed into a Mach--Zehnder interferometer with the detectors $D_1$ and $D_2$. The other photon $\gamma_i$ is directed into the delayed-choice measuring device with the detectors $D'_1$ and $D'_2$.}
\label{fig:SD_interferometer}
\end{figure}

\begin{figure}
\centering
    \includegraphics[width=0.95\textwidth]{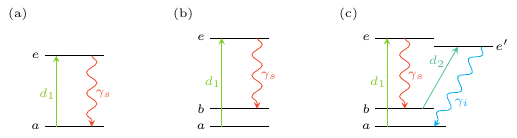}
\caption{Three different scenarios of the photon emission.
\textbf{(a)} The pulse $d_1$ excites $\ket{a}$ to $\ket{e}$. The state $\ket{e}$ returns back to $\ket{a}$, emitting a photon $\gamma_s$.
\textbf{(b)} The pulse $d_1$ excites $\ket{a}$ to $\ket{e}$. The state $\ket{e}$ is de-excited to $\ket{b}$, emitting a photon $\gamma_s$.
\textbf{(c)} A second pulse $d_2$ excites $\ket{b}$ to $\ket{e'}$. The state $\ket{e'}$ transits to $\ket{a}$, emitting a second photon $\gamma_i$.}
\label{fig:transition}
\end{figure}

The optical experiment of a Scully--Dr{\"u}hl-type quantum eraser is illustrated in \figref{fig:SD_interferometer}.
Two atoms (or other objects that can be triggered to emit photons) are located at $x$ on $\Path{1}$ and $y$ on $\Path{2}$. An incident photon pulse $d_1$ enters the beam splitter $\mathrm{BS_{in}}$, impinging and triggering either of the two atoms to emit a photon $\gamma_s$. The photon $\gamma_s$ travels along $\Path{1}$ and/or $\Path{2}$. Along $\Path{2}$, an adjustable phase shift $\theta$ is introduced. The two paths are recombined by the beam splitter $\mathrm{BS}_s$ before $\gamma_s$ clicks either of the two detectors, $D_1$ and $D_2$.\footnote{In \figref{fig:SD_interferometer}, we do not consider the polarization degree of freedom, and all beam splitters are non-polarizing ones, as opposed to the experiment in \figref{fig:interferometer}.}
We consider three different scenarios of how the atom emits $\gamma_s$ as depicted in \figref{fig:transition}.

\figref{fig:transition} (a) shows the case of two-level atoms. The incident pulse $d_1$ excites one of the two atoms from its initial state $\ket{a}$ to the excited state $\ket{e}$. The excited atom then emits a photon $\gamma_s$ while returns back to $\ket{a}$. Because both atoms end up in the same state $\ket{a}$, one cannot distinguish the which-way information about which atom emits $\gamma_s$. Correspondingly, the two-path interference is exhibited in the detection probabilities at $D_1$ and $D_2$.

\figref{fig:transition} (b) shows the case of three-level atoms. The pulse $d_1$ excites one of the two atoms from its initial state $\ket{a}$ to the excited state $\ket{e}$. The excited atom then emits a photon $\gamma_s$ while transits to a different state $\ket{b}$. The atom that emits $\gamma_s$ ends up in the state $\ket{b}$, while the other atom remains in the state $\ket{a}$. As the which-way information has been ``marked'' in terms of the states of the two atoms, $\Path{1}$ and $\Path{2}$ do not interfere with each other. Correspondingly, the detection probabilities at $D_1$ and $D_2$ does not exhibit any interference pattern.

\figref{fig:transition} (c) shows the same configuration as in (b) with an additional state $\ket{e'}$. After $\gamma_s$ is emitted from one of the two atoms, a second photon pulse $d_2$ is shot into $\mathrm{BS_{in}}$ to excite the atom in $\ket{b}$ to the state $\ket{e'}$. Subsequently, the atom in $\ket{e'}$ then emits a second photon $\gamma_i$ while returns back to the state $\ket{a}$. The which-way information recorded in one of the atoms is transferred to $\gamma_i$.
We then direct the photon $\gamma_i$ into the other Mach--Zehnder interferometer with a nonsymmetric beam splitter $\mathrm{BS}_i$ and the two detectors $D'_1$ and $D'_2$. Up to some phase factors that can be absorbed into $\theta$, the transformation matrix of $\mathrm{BS}_i$ can be specified by the unitary matrix \eqref{Ry}; correspondingly, the transmission and reflection coefficients are given by $\cos^2(\phi/2)$ and $\sin^2(\phi/2)$ respectively.
If we set $\phi=0$ or $\phi=\pi$ (i.e., $\mathrm{BS}_i$ is completely transparent or reflective), the which-way information of $\gamma_s$ can be inferred from whether $\gamma_i$ clicks $D'_1$ or $D'_2$. On the other hand, if we set $\phi=\pi/2$ (i.e., $\mathrm{BS}_i$ becomes symmetric), the which-way information is completely ``erased'', and correspondingly $\gamma_s$ is said to travel \emph{both} $\Path{1}$ and $\Path{2}$.

Consequently, within each subensemble associated with $D'_1$ or $D'_2$, the detection probabilities of $D_1$ and $D_2$ exhibit the interference between $\Path{1}$ and $\Path{2}$.
Furthermore, if $\phi$ is set to some value different from $0$, $\pi$, or $\pi/2$, the which-way information of each individual $\gamma_s$ is partially erased. In this case, within each subensemble associated with $D'_1$ or $D'_2$, the detection probabilities of $D_1$ and $D_2$ appear partially modulated in response to $\phi_1$, partially manifesting the two-path interference.
The distance from $x$ and $y$ to $\mathrm{BS}_i$ can be made longer than the lengths of $\Path{1}$ and $\Path{2}$ so that the value of $\phi$ can be adjusted in the delayed-choice manner.

The Scully--Dr{\"u}hl-type quantum eraser in \figref{fig:SD_interferometer} is \emph{formally} equivalent to the entanglement quantum eraser in \figref{fig:interferometer}. However, as opposed to the entanglement quantum eraser, the which-way information of $\gamma_s$ deduced from whether $\gamma_i$ clicks $D'_1$ or $D'_2$ is not always counterfactual, because the two ``recorders'' are in \emph{direct contact} with the two paths. In the case that $\phi=0$, a signal registered in $D'_1$ or $D'_2$ implies the \emph{factual} conclusion that $\gamma_s$ travels solely along $\Path{1}$ or $\Path{2}$, respectively. Similarly, in the case that $\phi=\pi$, a signal registered in $D'_1$ or $D'_2$ implies the factual conclusion the other way around.
Unlike the entanglement quantum eraser, the Scully--Dr{\"u}hl-type quantum eraser does bear some factual significance. Therefore, it makes good sense to say that the which-way information is marked in the first place and can be erased later if $\phi$ is adjusted to some values other than $0$ or $\pi$. The Scully--Dr{\"u}hl-type quantum eraser does present a ``mystery'' deeper than the standard EPR paradox.

In the optical double-slit experiment conducted by Kim \textit{et al.}\ \cite{PhysRevLett.84.1}, nonlinear optical crystal, BBO ($\beta\text{-BaB}_2\text{O}_4$), is placed at the two slits (corresponding to $x$ and $y$ in \figref{fig:SD_interferometer}) to generate signal-idler photon pairs via spontaneous parametric down-conversion (SPDC).
The finite width of the SPDC slit (approximately 0.3\,mm) introduces additional decoherence, blurring the nodes and antinodes, which leads to a significant reduction in the contrast of the recovered interference pattern compared to the optimal case. In contrast, as will be demonstrated later, the experiments conducted on both the IBM Quantum and IonQ platforms do not encounter this issue, allowing for nearly complete recovery of the interference pattern with clearly defined nodes and antinodes.

Furthermore, to truely demonstrate the retrocausal effect, the distances from $x$ and $y$ to $\mathrm{BS}_i$ must exceed those of $\Path{1}$ and $\Path{2}$, and the phase $\phi$ must be adjusted (either randomly or intentionally) in a delayed-choice manner, i.e., after the $D_{1/2}$ measurement has been performed. This presents a significant technical challenge in optical experiments, requiring both high agility and precision timing to modify the properties of $\mathrm{BS}_i$ within a brief window following the $D_{1/2}$ measurement but before the $D'_{1/2}$ measurement. The technique described in \cite{jacques2007experimental} for realizing Wheeler's delayed-choice experiment may be employed to overcome this challenge, using an electro-optical modulator (EOM) driven by a quantum random number generator (QRNG) to effectively render $\mathrm{BS}_i$ either completely transparent or symmetric with precision timing.

In the experiment by Kim \textit{et al.}\ \cite{PhysRevLett.84.1}, $\mathrm{BS}_i$ is not altered but remains a fixed symmetric beam splitter (referred to as BS in \cite{PhysRevLett.84.1}). To implement the delayed choice, two additional beam splitters (referred to as BSA and BSB) are placed along the path of $\gamma_i$ to either direct $\gamma_i$ to $D'_{1/2}$ (referred to as $\mathrm{D}_1$ and $\mathrm{D}_2$) or deflect it to a different pair of detectors (referred to as $\mathrm{D}_3$ and $\mathrm{D}_4$), with equal probability. As long as the distances from $x$ to BSA and from $y$ to BSB exceed those of $\Path{1}$ and $\Path{2}$, the random choice made by BSA and BSB can be considered a delayed-choice decision.
However, since the beam splitters BSA and BSB are continuously present, this introduces a philosophical loophole in retrocausality: one could argue that no retrocausal effect can be claimed, as everything could be explained deterministically, given that BSA and BSB are prearranged.\footnote{Similarly, the same philosophical loophole also applies to the experimental realization of the entanglement quantum eraser performed by Walborn \textit{et al.}\ \cite{PhysRevA.65.033818}, where, correspondingly, the value of $\phi$ in \figref{fig:interferometer} is fixed in advance.}

In contrast, on the IBM Quantum platform, \emph{delay gates} can be applied to postpone the (random or intentional) adjustment of $\phi$, ensuring that the choice is made in a delayed-choice manner. Since gate operations are executed sequentially in time, any involvement of the choice is entirely absent before the signal qubit is measured, thereby closing the aforementioned loophole. Even if the EOM technique is adopted in optical experiments to alter $\mathrm{BS}_i$ with precision timing, the IBM Quantum experiments still hold an advantage in that $\phi$ can be precisely set to any desired value, whereas the EOM method typically only allows $\mathrm{BS}_i$ to be rendered either fully transparent or symmetric.

\subsection{Remarks on conceptual and experimental issues}
We have emphasized the distinction between the entanglement quantum eraser and the Scully--Dr{\"u}hl-type quantum eraser. However, contrary to what might be inferred, this paper does not primarily focus on philosophical issues, which are addressed only as motivational context. Instead, our investigation is centered on the physical and experimental aspects. For a more in-depth philosophical discussion, we refer interested readers to \cite{chiou2023delayed}.

The distinction between these two types of quantum erasers is physical, not merely
philosophical. From the perspective of testing the foundations of quantum mechanics, this distinction is fundamental and significant. It is crucial to experimentally investigate both types, as a more fundamental theory beyond standard quantum mechanics could validate the predictions of quantum mechanics for one type while potentially invalidating them for the other.

As discussed earlier, neither the optical experiment for the entanglement quantum eraser \cite{PhysRevA.65.033818} nor that for the Scully--Dr{\"u}hl-type quantum eraser \cite{PhysRevLett.84.1} has realized erasure in a genuinely delayed-choice manner. Therefore, as tests of the foundations of quantum mechanics, these experiments still have significant room for improvement (recall Footnote~\ref{foot:loophole}), such as incorporating the challenging EOM techniques used in \cite{jacques2007experimental}.

By contrast, as will be demonstrated, implementing the Scully--Dr{\"u}hl-type quantum eraser in quantum circuits not only offers various advantages over optical experiments but, more importantly, enables a truly random choice in a genuinely delayed-choice manner.\footnote{Likewise, implementing the entanglement quantum eraser in quantum circuits exhibits similar advantages, as demonstrated in \cite{chiou2024complementarity}. However, since the primary focus of \cite{chiou2024complementarity} is on issues related to complementarity relations, its experiments on the entanglement quantum eraser have not achieved a genuinely delayed choice with the same level of rigor as those for the Scully--Dr{\"u}hl-type quantum eraser demonstrated in this paper---although, in principle, this could be achieved.}

As a significant contribution to testing the foundations of quantum mechanics, this work demonstrates quantum erasers with essential improvements that have not been realized in any previous experimental setting. The significance of this work extends beyond merely implementing quantum erasers in quantum circuits.

Furthermore, it is important to acknowledge the idealizations made in illustrating the ideas in \secref{sec:SD quantum eraser}. For simplicity, in the process of photon absorption or emission in \figref{fig:transition}, we have disregarded any footprints left on external systems, as well as the remaining degrees of freedom of the atoms (e.g., recoil) other than the energy states.

Since photon absorption by an atom is inherently nonunitary when all degrees of freedom are taken into account \cite{zhang2024examples}, one might argue that when the driving pulse interacts with the two atoms $x$ and $y$ in \figref{fig:SD_interferometer}, it is unambiguously absorbed by only one of them, thereby preventing two-path interference. However, nonunitarity is a matter of degree and does not
necessarily preclude the possibility of two-path interference. If the atoms are sufficiently massive, recoil effects become negligible, allowing interference to occur. This has been experimentally demonstrated in a double-slit-like setup---first by Eichmann \textit{et al.}\ \cite{PhysRevLett.70.2359, PhysRevA.57.4176}, who used a laser beam to excite two trapped ${}^{198}\mathrm{Hg}^+$ ions, and more recently by Araneda \textit{et al.}\ \cite{PhysRevLett.120.193603}, who used two trapped ${}^{138}\mathrm{Ba}^+$ ions.
In both cases, the observed interference patterns exhibited significantly lower visibility than the idealized prediction. This reduction in visibility can be attributed to unavoidable interactions with the atoms' remaining degrees of freedom, as well as with external systems and fields \cite{PhysRevResearch.2.012031}.

A similar phenomenon occurs in optical experiments without trapped atoms. In \cite{PhysRevLett.84.1}, the finite width of the SPDC slit introduces additional degrees of freedom, inevitably reducing the visibility of interference patterns, as discussed earlier. For further examples and analyses of decoherence in two-path interference due to additional degrees of freedom, we refer readers to Section 4 of \cite{chiou2023delayed}.

By comparison, as will be demonstrated, experiments performed in quantum circuits can, in principle, achieve nearly ideal interference patterns, closely matching the idealized prediction. This is made possible by the high fidelity of state-of-the-art quantum circuit technology, representing a significant advantage of conducting quantum experiments in such systems.

\section{Implementation in quantum circuits}\label{sec:quantum circuit}
The entanglement quantum eraser in \figref{fig:interferometer} can be implemented in a quantum circuits as studied in detail in \cite{chiou2024complementarity}.

The Scully--Dr{\"u}hl-type delayed-choice quantum eraser experiment can also be implemented in a quantum circuit as shown in \figref{fig:quantum circuit simple}. The phase gate $P(\theta)$ is given by
\begin{equation}
P(\theta) \equiv e^{i\theta/2}R_z(\theta)
=
\left(
  \begin{array}{cc}
    1 & 0 \\
    0 & e^{i\theta} \\
  \end{array}
\right),
\end{equation}
and the $R_y(\phi)$ gate is given by
\begin{equation}\label{Ry}
R_y(\phi) \equiv e^{-i\phi Y/2}=\cos\frac{\phi}{2}I-i\sin\frac{\phi}{2}Y
=
\left(
  \begin{array}{cc}
    \cos\frac{\phi}{2} & -\sin\frac{\phi}{2} \\
    \sin\frac{\phi}{2} & \cos\frac{\phi}{2} \\
  \end{array}
\right).
\end{equation}
Analogous to $\mathrm{BS_{in}}$ in \figref{fig:SD_interferometer}, the first Hadamard ($H$) gate transforms the ``$s$'' qubit from the state $\ket{0}$ into $1/\sqrt{2}(\ket{0}+\ket{1})$. After this $H$ gate, the states $\ket{0}$ and $\ket{1}$ are to be viewed as $\Path{1}$ and $\Path{2}$ respectively by analogy. Analogous to the adjustable phase-shift plate, the $P(\theta)$ gate adds a relative phase $e^{i\theta}$ to $\ket{1}$. Analogous to $\mathrm{BS}_s$, the second (dahsed) Hadamard ($H$) gate recombines the two ``paths''. Finally, the $0/1$ readouts of $D_s$ is analogous to the signals registered in $D_1$ and $D_2$.
Meanwhile, the ``$i$'' qubit serves as the recorder of the which-way information. Through the CNOT gate, the ``$i$'' qubit makes direct contact with the ``$s$'' qubit and records its which-way information. The $R_y(\phi)$ gate is analogous to $\mathrm{BS}_i(\phi)$, and the $0/1$ readouts of $D_i$ is analogous to the signals registered in $D'_1$ and $D'_2$.
Furthermore, a delay gate might be applied to ensure that the value of $\phi$ is set in the delay-choice manner.


\begin{figure}
\begin{quantikz}
\lstick{$\ket{0}_s$} & \gate{H} & \ctrl{1} & \gate{P(\theta)} & \gate[style=dashed]{H}  & \meter{$D_s$} \\
\lstick{$\ket{0}_i$} & \qw  & \targ{}   & \qw & \qw & \gate{\text{Delay}}  & \gate{R_y(\phi)} & \meter{$D_x$}
\end{quantikz}
\caption{Implementation of a Scully--Dr{\"u}hl-type delayed-choice quantum eraser in a quantum circuit.}
\label{fig:quantum circuit simple}
\end{figure}

In the quantum circuit in \figref{fig:quantum circuit simple} the which-way information is recorded in a single qubit, whereas in the experiment in \figref{fig:SD_interferometer} the which-way information is recorded in two \emph{spatially separated} recorders. In a sense, the quantum eraser in \figref{fig:quantum circuit simple} is not genuinely of the Scully--Dr{\"u}hl type. To draw the analogy even closer, we instead consider the quantum circuit in \figref{fig:quantum circuit}. The ``$x$'' and ``$y$'' qubits serve as the two recorders of the which-way information, as the ``$\Path{1}$'' state of the ``$s$'' qubit flips the state of the ``$x$'' qubit, whereas the ``$\Path{2}$'' state flips the state of the ``$y$'' qubit. The the $R_y(\phi)$ gate together with the CNOT gate prior to it is analogous to $\mathrm{BS}_i(\phi)$. The two recorders both make direct contact with the ``$s$'' qubit and remain spatially separated from each other. They do not interact with each other until the CNOT gate before the $R_y(\phi)$ gate is applied on them.

\begin{figure}
\begin{quantikz}
\lstick{$\ket{0}_s$} & \gate{H} & \octrl{1} & \ctrl{2} & \gate{P(\theta)} & \gate[style=dashed]{H}  & \meter{$D_s$} \\
\lstick{$\ket{0}_x$} & \qw  & \targ{} & \qw     & \qw & \qw & \gate{\text{Delay}} & \ctrl{1} & \gate{R_y(\phi)} & \meter{$D_x$} \\
\lstick{$\ket{0}_y$} & \qw  & \qw     & \targ{} & \qw & \qw & \gate{\text{Delay}} & \targ{}  & \qw              & \meter{$D_y$}
\end{quantikz}
\caption{Implementation of a Scully--Dr{\"u}hl-type delayed-choice quantum eraser in a quantum circuit with two ``recorders'' of the which-way information.}
\label{fig:quantum circuit}
\end{figure}

We can further connect the $R_y(\phi)$ gate with an additional qubit $\ket{q}_a$ as shown in \figref{fig:random choice}. This enables the circuit to automatically make random choice with equal probability between setting $\phi$ to a nonzero value and setting $\phi$ to zero.
Replacing the $R_y(\phi)$ gate with a controlled version, $CR_y(\phi)$, enhances the retrocausal effect by making the choice of erasure both delayed and random, in the same manner as using an EOM driven by a QRNG in the work of \cite{jacques2007experimental}, as discussed in \secref{sec:SD quantum eraser}.
When the readout of $D_a$ yields 1, it amounts to setting a nonzero $\phi$; when the readout of $D_a$ yields 0, it amounts to setting $\phi=0$.

\begin{figure}
\begin{quantikz}
& \lstick{$\cdots$}  & \gate{R_y(\phi)} & \qw & \lstick{$\cdots$} \\
\lstick{$\ket{0}_a$} & \gate{H} & \ctrl{-1} & \meter{$D_a$}
\end{quantikz}
\caption{The circuit component for the random choice of applying the $R_y(\phi)$ gate.}
\label{fig:random choice}
\end{figure}

\begin{figure}
\begin{quantikz}
\lstick{$\ket{0}_s$} & \gate{H} \slice{1} & \octrl{1} & \ctrl{2} \slice{2} & \gate{P(\theta)} \slice{3} & \qw \slice{4} & \qw \slice{5} & \gate[style=dashed]{H} \slice{6} & \meter{$D_s$} \\
\lstick{$\ket{0}_x$} & \qw      & \targ{}  & \qw     & \qw & \ctrl{1} & \gate{R_y(\phi)} & \qw & \meter{$D_x$} \\
\lstick{$\ket{0}_y$} & \qw      & \qw      & \targ{} & \qw & \targ{}  & \qw              & \qw & \meter{$D_y$}
\end{quantikz}
\caption{Equivalent layout to \figref{fig:quantum circuit}.}
\label{fig:quantum circuit 2}
\end{figure}

To calculate the outcomes of the quantum circuit in \figref{fig:quantum circuit}, we consider the equivalent layout as shown in \figref{fig:quantum circuit 2} for convenience without worrying exactly when each gates is applied. The quantum state at each slice can be straightforwardly calculated as
\begin{subequations}
\begin{eqnarray}
\ket{\psi_1} &=& \frac{1}{\sqrt{2}}(\ket{0}+\ket{1})\ket{00},\\
\ket{\psi_2} &=& \frac{1}{\sqrt{2}}\big(\ket{0}\ket{10}+\ket{1}\ket{01}\big),\\
\ket{\psi_3} &=& \frac{1}{\sqrt{2}}\left(\ket{0}\ket{10}+e^{i\theta}\ket{1}\ket{01}\right),\\
\ket{\psi_4} &=& \frac{1}{\sqrt{2}}\left(\ket{0}\ket{11}+e^{i\theta}\ket{1}\ket{01}\right),\\
\ket{\psi_5} &=& \frac{1}{\sqrt{2}}\ket{0}
\left(-\sin\frac{\phi}{2}\ket{0}+\cos\frac{\phi}{2}\ket{1}\right)\ket{1} + \frac{e^{i\theta}}{\sqrt{2}}\ket{1}
\left(\cos\frac{\phi}{2}\ket{0}+\sin\frac{\phi}{2}\ket{1}\right)\ket{1},\\
\label{psi6}
\ket{\psi_6} &=& \frac{1}{2}(\ket{0}+\ket{1})
\left(-\sin\frac{\phi}{2}\ket{0}+\cos\frac{\phi}{2}\ket{1}\right)\ket{1} + \frac{e^{i\theta}}{2}(\ket{0}-\ket{1})
\left(\cos\frac{\phi}{2}\ket{0}+\sin\frac{\phi}{2}\ket{1}\right)\ket{1} \nonumber\\
&\equiv&
\frac{1}{2}\left[
\left(\cos\frac{\phi}{2}+e^{i\theta}\sin\frac{\phi}{2}\right)\ket{0}
+ \left(\cos\frac{\phi}{2}-e^{i\theta}\sin\frac{\phi}{2}\right)\ket{1}
\right] \ket{11}\nonumber\\
&& \mbox{} -\frac{1}{2}\left[
\left(\sin\frac{\phi}{2}-e^{i\theta}\cos\frac{\phi}{2}\right)\ket{0}
+ \left(\sin\frac{\phi}{2}+e^{i\theta}\cos\frac{\phi}{2}\right)\ket{1}
\right] \ket{01} .
\end{eqnarray}
\end{subequations}

If we focus on the ``$s$'' qubit $\ket{q}_s$, it is described by the reduced density matrix $\rho^{(s)}_n:= \Tr_{\ket{q}_x,\ket{q}_y}(\ket{\psi_n}\bra{\psi_n})$ traced out over the ``$x$'' and ``$y$'' qubits $\ket{q}_x$ and $\ket{q}_y$. Particularly, the reduced density matrices at slices 5 and 6 are
\begin{equation}
\rho^{(s)}_5 = \rho^{(s)}_6 = \frac{1}{2}\big(\ket{0_s}\bra{0_s} + \ket{1_s}\bra{1_s}\big).
\end{equation}
Both in the ``closed'' configuration (i.e., the dashed $H$ gate is present) and the ``open'' configuration (i.e., the dashed $H$ gate is removed), the probabilities that $D_s$ yields the outcomes $0$ and $1$ are given by
\begin{subequations}\label{p s}
\begin{eqnarray}
  p(0_s) &=& \Tr \left(\ket{0_s}\bra{0_s} \rho^{(s)}_{5,6}\right) = \frac{1}{2},\\
  p(1_s) &=& \Tr \left(\ket{1_s}\bra{1_s} \rho^{(s)}_{5,6}\right) = \frac{1}{2}.
\end{eqnarray}
\end{subequations}
The outcomes of $D_s$ exhibit no interference between the two ``paths'' $\ket{0}$ and $\ket{1}$ (i.e., independent of the phase shift $\theta$) whether in the closed or open configurations.

On the other hand, if we focus on the subsystem of the ``$x$'' and ``$y$'' qubits, it is described by the reduced density matrix $\rho^{(xy)}_n:= \Tr_{\ket{q}_s}(\ket{\psi_n}\bra{\psi_n})$ traced out over the ``$s$'' qubit. Particularly, the reduced density matrices at slice 5 and 6 are
\begin{equation}
\rho^{(xy)}_5 = \rho^{(xy)}_6 = \frac{1}{2}\big(\ket{1_x1_y}\bra{1_x1_y} + \ket{0_x1_y}\bra{0_x1_y}\big).
\end{equation}
Both in the closed and open configurations, $D_x$ and $D_y$ yield only the two possible outcomes $0_x1_y$ and $1_x1_y$ with equal probabilities, i.e.,
\begin{subequations}\label{p xy}
\begin{eqnarray}
  p(1_x1_y) &=& \Tr \left(\ket{1_x1_y}\bra{1_x1_y} \rho^{(xy)}_{5,6}\right) = \frac{1}{2},\\
  p(0_x1_y) &=& \Tr \left(\ket{0_x1_y}\bra{0_x1_y} \rho^{(xy)}_{5,6}\right) = \frac{1}{2},\\
  \label{p xy leaked}
  p(0_x0_y) &=& p(1_x0_y) = 0.
\end{eqnarray}
\end{subequations}
The outcomes $1_x1_y$ and $0_x1_y$ of $D_x$ and $D_y$ are analogous to the signals registered in $D'_1$ and $D'_2$ respectively in \figref{fig:SD_interferometer}.

Theoretically, the outcomes $0_x0_y$ and $1_x0_y$ should never occur, as indicated in \eqref{p xy leaked}. However, in real experiments, due to noise and errors in quantum circuits, these outcomes still appear as ``leakage''. This leakage is analogous to the ``dark counts'' and ``false counts'' in optical experiments, which arise from path loss, beam splitter loss, and imperfect photon detection efficiency.
In our analysis of experimental results, we simply discard these leaked outcomes. The characteristics of leakage in the IBM Quantum and IonQ platforms are provided as calibration data in Appendix~\ref{app:calibration}.

\subsection{Which-way information}
The which-way information of $\ket{q}_s$ can be explicitly measured in the open configuration, where the two ``paths'' $\ket{0}$ and $\ket{1}$ register the readouts 0 and 1 separately in $D_s$, and thus the readout of $D_s$ unambiguously determines the which-way information of $\ket{q}_s$.
In the closed configuration, on the other hand, the which-way information cannot be determined from the readout of $D_s$, but it can be indirectly inferred with a certain degree of certainty from the paired readout of $D_x$ and $D_y$. Adopt the guessing strategy as follows: the which-way information of $\ket{q}_s$ is guessed to be $\ket{0}$ if $D_x$ and $D_y$ yield $1_x1_y$, and $\ket{1}$ if $D_x$ and $D_y$ yield $0_x1_y$. The probability of successfully guessing the which-way information can be empirically computed from the concurrence counts between the outcomes of $D_s$ and the pair of $D_x$ and $D_y$ in the open configuration.
Mathematically, the probability of success is computed as
\begin{eqnarray}\label{p succ}
p_\mathrm{succ} &=& p(1_x1_y)\,p(0_s|1_x1_y) + p(0_x1_y)\,p(1_s|0_x1_y)
\equiv p(0_s1_x1_y) + p(1_s0_x1_y)\nonumber\\
&=&
\left|\langle0_s1_x1_y\ket{\psi_5}\right|^2
+
\left|\langle1_s0_x1_y\ket{\psi_5}\right|^2
\nonumber\\
&=& \frac{1+\cos\phi}{2}.
\end{eqnarray}
The \emph{distinguishability} of the which-way information is defined as
\begin{equation}\label{D def}
\mathcal{D}:= 2\,p_\mathrm{succ}-1,
\end{equation}
the absolute value of which quantifies how much the which-way information can be deduced based on the outcomes of $D'_1$ and $D'_2$.
(In case that $\mathcal{D}<0$, it simply means that the strategy should have been the other way around.)
By \eqref{p succ}, we have
\begin{equation}\label{D}
\mathcal{D}=\cos\phi.
\end{equation}

\subsection{Interference pattern recovered}
In the closed configuration, in regard to the outcomes of $D_x$ and $D_y$, we can consider the subensemble of the events associated with the readout $1_x1_y$ and the subensemble associated with the readout $0_x1_y$ \emph{separately}. Within either of the two subensembles (labeled with ``$1_x1_y$'' and ``$0_x1_y$'' respectively), the which-way information of $\ket{q}_s$ can be partially or completely erased. Accordingly, the interference pattern of $D_s$ within the confines of either subensemble can be partially or completely recovered.

According to \eqref{psi6}, for the events corresponding to $1_x1_y$, the wavefunction of $\ket{q}_s$ is collapsed into
\begin{equation}
\ket{\psi} = \frac{1}{\sqrt{2}}\left[
\left(\cos\frac{\phi}{2}+e^{i\theta}\sin\frac{\phi}{2}\right)\ket{0}
+ \left(\cos\frac{\phi}{2}-e^{i\theta}\sin\frac{\phi}{2}\right)\ket{1}
\right].
\end{equation}
Within the $1_x1_y$ subensemble, the probabilities of having the readout 0 and the readout 1 in $D_s$ and are given respectively by
\begin{subequations}\label{p 1x1y}
\begin{eqnarray}
p(0_s|1_x1_y) &=& \frac{\big|\langle0\ket{\psi}\big|^2}{\big|\langle\psi\ket{\psi}\big|^2}
= \frac{1}{2}(1 + \sin\phi\cos\theta), \\
p(1_s|1_x1_y) &=& \frac{\big|\langle0\ket{\psi}\big|^2}{\big|\langle\psi\ket{\psi}\big|^2}
= \frac{1}{2}(1 - \sin\phi\cos\theta).
\end{eqnarray}
\end{subequations}
Similarly, for the events corresponding to $0_x1_y$, the wavefunction of $\ket{q}_s$ is collapsed into
\begin{equation}
\ket{\psi} = -\frac{1}{\sqrt{2}}\left[
\left(\sin\frac{\phi}{2}-e^{i\theta}\cos\frac{\phi}{2}\right)\ket{0}
+ \left(\sin\frac{\phi}{2}+e^{i\theta}\cos\frac{\phi}{2}\right)\ket{1}
\right].
\end{equation}
Within the $0_x1_y$ subensemble, the probabilities of having the readout 0 and the readout 1 in $D_s$ and are given respectively by
\begin{subequations}\label{p 0x1y}
\begin{eqnarray}
p(0_s|0_x1_y) &=& \frac{\big|\langle0\ket{\psi}\big|^2}{\big|\langle\psi\ket{\psi}\big|^2}
= \frac{1}{2}(1 - \sin\phi\cos\theta), \\
p(1_s|0_x1_y) &=& \frac{\big|\langle0\ket{\psi}\big|^2}{\big|\langle\psi\ket{\psi}\big|^2}
= \frac{1}{2}(1 + \sin\phi\cos\theta).
\end{eqnarray}
\end{subequations}

The \emph{visibility} of the interference pattern within either of the subensembles is defined as
\begin{equation}\label{V def}
\mathcal{V} := \frac{\max_\theta p(0_i|\dots)-\min_\theta p(0_i|\dots)}
{\max_\theta p(0_i|\dots)+\min_\theta p(0_i|\dots)},
\end{equation}
where ``$\dots$'' represents either $1_x1_y$ or $0_x1_y$.
By \eqref{p 1x1y} and \eqref{p 0x1y}, we have
\begin{equation}\label{V}
\mathcal{V} = \left|\sin\phi\right|.
\end{equation}

The distinguishability given by \eqref{D} and the visibility given by \eqref{V} affirm the complementarity relation: the more the which-way information can be deduced, the less the interference pattern is recovered.
In fact, \eqref{D} and \eqref{V} saturate the wave--particle duality relation \cite{PhysRevA.51.54, PhysRevLett.77.2154}:
\begin{equation}
\mathcal{V}^2+\mathcal{D}^2 \leq 1.
\end{equation}
Because the quantum circuit in \figref{fig:quantum circuit} for the Scully--Dr{\"u}hl-type quantum eraser shares exactly the same mathematical structure with the quantum circuit for the entanglement quantum eraser considered in \cite{chiou2024complementarity},\footnote{Note that \cite{chiou2024complementarity} also considers the extension that the degree of entanglement between the signal and idler quantons/qubits is adjustable. This extension has no direct analogy in the Scully--Dr{\"u}hl-type quantum eraser and is not considered in this paper.} the meaning and significance of the complementarity relations between interference visibility and which-way distinguishability can be understood in the same way as discussed in \cite{chiou2024complementarity} (also see the references thereof).

\subsection{Remarks on ``two-path'' interference}
In many experimental realizations and proposals of the entanglement quantum eraser and the Scully--Dr\"{u}hl-type quantum eraser (see \cite{RevModPhys.88.015005}), interference occurs between different quantum states, which do not necessarily correspond to a single quanton localized in distinct spatial paths or locations. In such cases, the term ``two-path'' interference is more accurately described as ``two-state'' interference, and ``which-way'' information should be referred to as ``which-state'' information. Nevertheless, the conventional terminology of ``two-path'' and ``which-way'' remains widely used in the literature, and we adhere to this nomenclature for consistency.

In the quantum circuits depicted in \figref{fig:quantum circuit simple} and \figref{fig:quantum circuit}, the interference occurs between the two states $\ket{0}_s$ and $\ket{1}_s$. In some qubit architectures, such as a single photon in the dual-rail representation for optical quantum computers \cite{PhysRevA.52.3489} and a single electron in a double quantum dot (DQD) \cite{PhysRevLett.105.246804, PhysRevLett.95.090502, PhysRevB.84.161302}, these states indeed correspond to different spatial localizations of a single quanton. However, in superconducting transmon architectures, such as those employed in the IBM Quantum platform, and in trapped ion architectures, such as the IonQ platform, $\ket{0}_s$ and $\ket{1}_s$ do not correspond to different spatial positions of a single quanton.

The essence of retrocausality in quantum erasure is that which-state information of a quantum state can be initially marked and subsequently erased in a delayed-choice manner. Importantly, whether the quantum state corresponds to a spatially localized quanton is not essential to this phenomenon. Thus, our experiments conducted on the IBM Quantum and IonQ platforms are genuine quantum eraser experiments, not to be mistaken for mere computational simulations on quantum computers.

Under extreme conditions---perhaps requiring the free will of a conscious being---fundamental principles beyond standard quantum mechanics may sustain the retrocausal effect when the quantum state lacks spatial localization, while potentially restricting it in cases where spatial localization is prominent.
Although highly unlikely, this possibility remains conceivable because spatial positions play a special role in decoherence theory and the many-worlds interpretation (MWI) of quantum mechanics. In decoherence theory, the position basis of a macroscopic object serves as a preferred (or \emph{pointer}) basis, as system--environment interactions typically involve position-dependent terms that lead to rapid localization, whereas decoherence of other degrees of freedom occurs on much longer timescales \cite{PhysRevD.26.1862, RevModPhys.75.715, schlosshauer2007quantum, DieterZeh2007}. In MWI, decoherence leads to \emph{einselection} (i.e., environment-induced superselection), which typically selects spatial positions as the preferred basis, thereby giving rise to spatial branching in the many-worlds picture \cite{PhysRevD.26.1862, RevModPhys.75.715, wallace2012emergent}.

In the IBM Quantum and IonQ platforms, the quantum states undergoing interference do not correspond to distinct localizations of a single quanton. To nevertheless manifest spatial localization in our setup, we have modified \figref{fig:quantum circuit simple} into \figref{fig:quantum circuit} by introducing two spatially separated recorders. The fact that information recorded in two distinct locations can still be erased retroactively is significant. Whether spatial positions play a fundamental role in quantum erasers warrants further experimental investigation under different settings and conditions. This work opens a new avenue for investigating quantum erasure in quantum circuits and motivates further experiments across diverse qubit architectures to deepen our understanding of its underlying principles.

\section{Experiments on IBM Quantum}\label{sec:IBM}
\begin{figure}
\centering
\begin{picture}(450,212)(0,0)
\put(20,30){\includegraphics[width=0.26\textwidth]{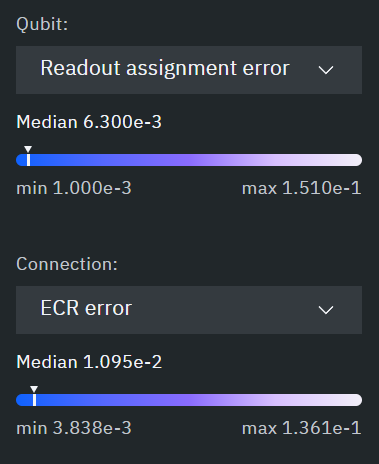}}
\put(168,0){\includegraphics[width=0.56\textwidth]{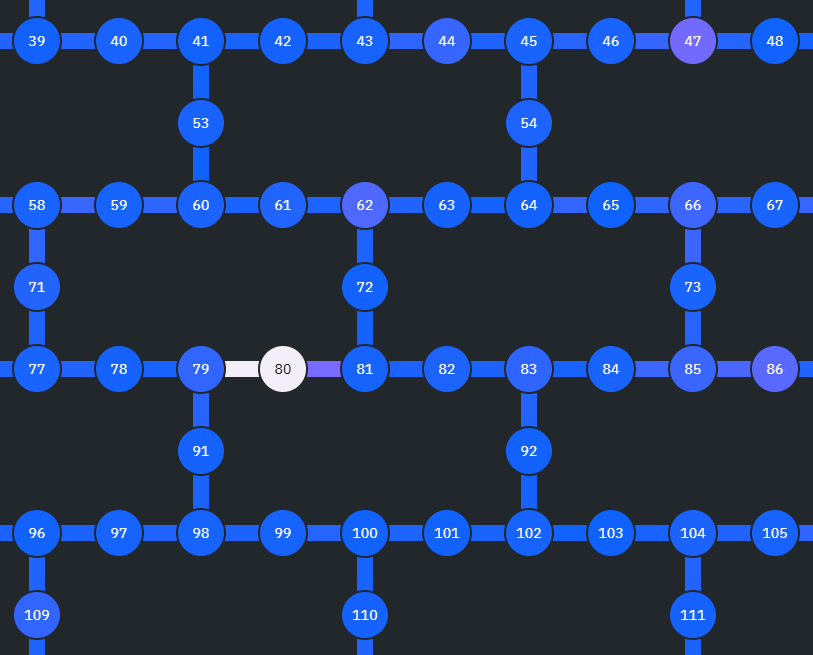}}
\end{picture}
\caption{Portion of the qubit layout for the IBM Quantum processor \texttt{ibm\_kyiv}, showing the qubits used in the experiments.
The numbers index the physical qubits and the links denote the connections that allow CNOT operations (via ECR operations).}\label{fig:layout}
\end{figure}

In this section, we present the experimental results performed in the quantum processor of IBM Quantum based on the superconducting transmon architecture \cite{IBMQ}.
These experiments are conducted in the processor \texttt{ibm\_kyiv}, which has 127 qubits in a chip as depicted in \figref{fig:layout}.\footnote{\figref{fig:layout} is obtained from the IBM Quantum website \cite{IBMQ}.} The distance between neighboring qubits is on the order of $10^2\mu\,\text{m}$ \cite{chow2015characterizing}. This ensures that the recorders marking the which-way information remain spatially separated by at least $\sim10^2\mu$m. Since \texttt{ibm\_kyiv} is not fully connected via the echoed cross-resonance (ECR) gate (which is equivalent to a CNOT gate up to single-qubit pre-rotations) between all pairs of qubits, we must invoke the SWAP operation (which can be decomposed into three CNOT gates) somewhere to realize all the two-qubit gates in \figref{fig:quantum circuit} and \figref{fig:random choice}.

Specifically, we choose the qubits 40, 41, 42, and 53 from \figref{fig:layout} to perform experiments, as they are close to one another and have relatively low readout and ECR errors. The detailed implementation on the physical qubits is illustrated in \figref{fig:transpiled}. The subscript $i$ of $\ket{q}_i$ indicates the number index of the physical qubits in \figref{fig:layout}.
The main component in \figref{fig:quantum circuit 2} and the random choice component in \figref{fig:random choice} are combined into the realistic layout in \figref{fig:transpiled}, initially with the qubit mapping $(s,x,y,a)\mapsto (41,42,53,40)$.
After the SWAP gate in \figref{fig:transpiled}, $\ket{q}_{41}$ and $\ket{q}_{42}$ interchanges their roles, leading to the new qubit mapping $(s,x,y,a)\mapsto (42,41,53,40)$.

If the measurement $D_s$ for the state of $\ket{q}_s$ is performed in the end (i.e., at the moment indicated by ``trash'' in \figref{fig:transpiled}), the random choice invoked by the $CR_y(\phi)$, i.e., controlled-$R_y(\phi)$, gate is made before $D_s$. That is, the which-way information can be partially or completely erased by $CR_y(\phi)$, but the erasure is not made in the delayed-choice manner.\footnote{\label{foot:delayed choice}However, the exact moment when the erasure occurs is open to interpretation. One perspective is that the random choice (and thus the erasure) is made when the $CR_y(\phi)$ gate is invoked. Alternatively, it could be argued that the choice is not finalized until the state of the ancilla qubit is known, which occurs when the measurement $D_a$ is performed. According to this latter view, one could still assert that the erasure is achieved in a delayed-choice manner, provided that $D_a$ does not causally precede $D_s$ (e.g., if $D_s$ and $D_a$ are measured simultaneously). For further discussion on the interpretative issues, see \cite{chiou2023delayed}.}


\begin{figure}
\begin{quantikz}
\lstick{$\ket{0}_{40}$}&\qw&\qw&\qw&\qw&\qw&\qw&\gate[style=dashed]{\text{Delay}}&\gate{H}&\qw& \ctrl{1}&\qw& \meter{$D_a$}\\
\lstick{$\ket{0}_{41}$} & \gate{H} & \ctrl{2} & \octrl{1} & \gate{P(\theta)} & \gate{H}  & \meter{$D_s$}&\gate{\text{Delay}} & \swap{1}&\ctrl{2} & \gate{R_y(\phi)}&\qw & \meter{$D_x$}\\
\lstick{$\ket{0}_{42}$} & \qw  &\qw   &\targ{}     & \qw & \qw& \qw&\gate[style=dashed]{\text{Delay}}&\swap{-1} & \qw & \qw & \qw & \trash{\text{trash}} \\
\lstick{$\ket{0}_{53}$} & \qw& \targ{}  & \qw      & \qw & \qw &\qw &\gate[style=dashed]{\text{Delay}}&\qw & \targ{}&\qw & \qw  & \meter{$D_y$}
\end{quantikz}
\caption{The implementation of the Scully--Dr{\"u}hl-type delayed-choice quantum eraser with a specific qubit mapping on \texttt{ibm\_kyiv}.}
\label{fig:transpiled}
\end{figure}

The architecture of IBM Quantum allows the measurement $D_s$ to be executed midway, as illustrated in \figref{fig:transpiled}. This ensures that the random choice made by the $CR_y(\phi)$ gate is truly performed in a delayed-choice manner---that is, the $CR_y(\phi)$ operation is only invoked after the outcome of $D_s$ has been determined. Additionally, a \emph{delay gate} providing an adjustable delay time $t_\text{delay}$ can be applied immediately after the measurement $D_s$ to further extend the delay before the random choice is made.\footnote{Despite being called a ``gate'', the delay gate does nothing to the qubit at all; it merely enforces a pause in the execution schedule for a specified duration before the next gate is applied. This can be observed in the exact schedules depicted in \figref{fig:low level 2options delay5000} and \figref{fig:low level 4options delay5000}.} Since gate operations are executed sequentially in time, all gates to the right of the delay gates in \figref{fig:transpiled} are entirely absent until the specified delay time has elapsed.\footnote{For a given $t_\text{delay}$, a delay gate with duration $t_\text{delay}$ is inserted at $\ket{q}_{41}$, as shown in \figref{fig:transpiled}. The transpiler then adjusts the delay times for other gates (indicated by dashed lines) to ensure synchronized gate operations. Refer to \figref{fig:low level 2options delay5000} for the transpiled low-level circuit in terms of primitive gates along with the exact gate execution schedule.}
In the following, we present the experimental results with various values of $t_\text{delay}=0$.

\begin{figure}
\centering
    \includegraphics[width=0.63\textwidth, angle = 90]{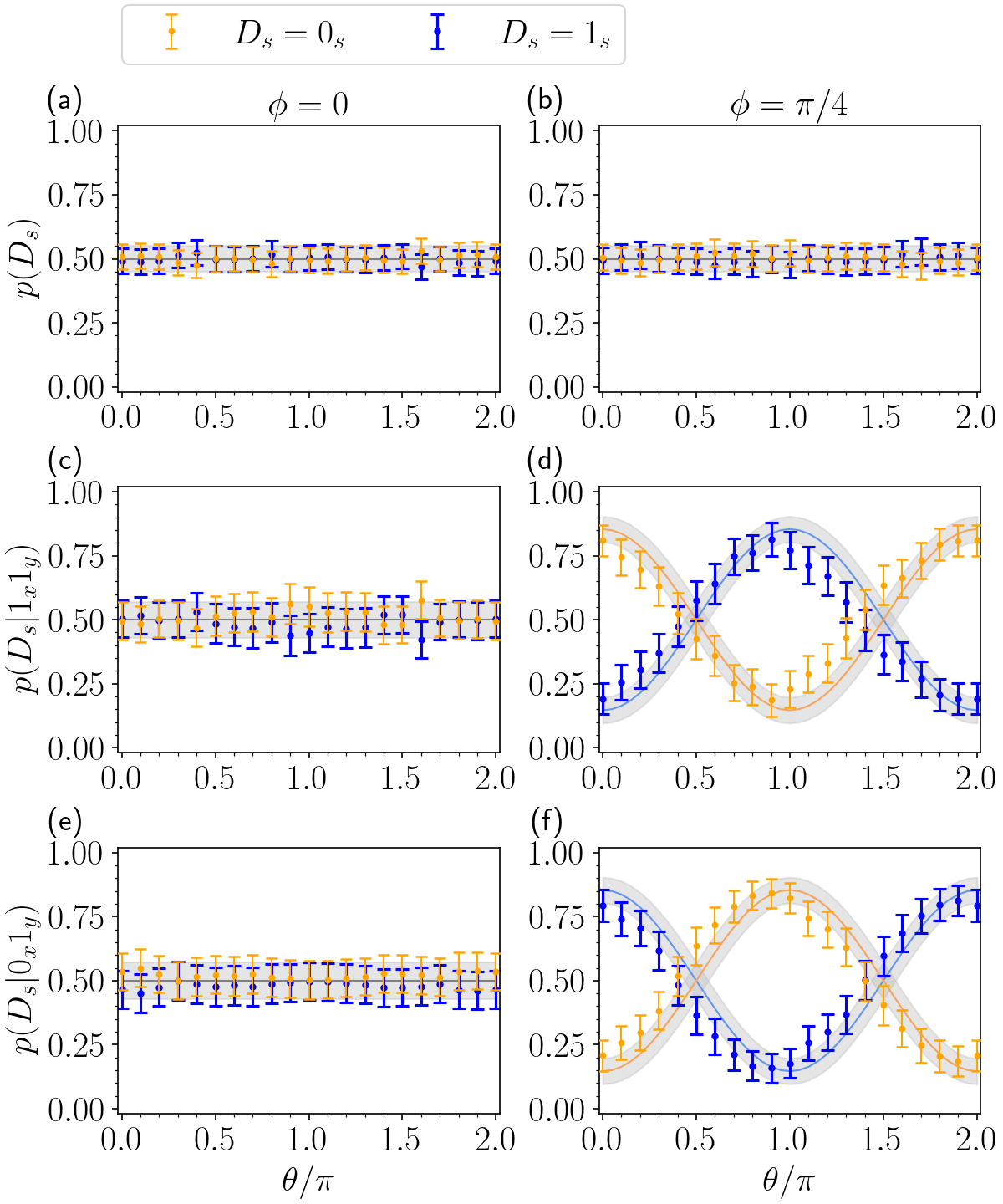}
    \vspace{0.6cm}
    \includegraphics[width=0.63\textwidth, angle = 90]{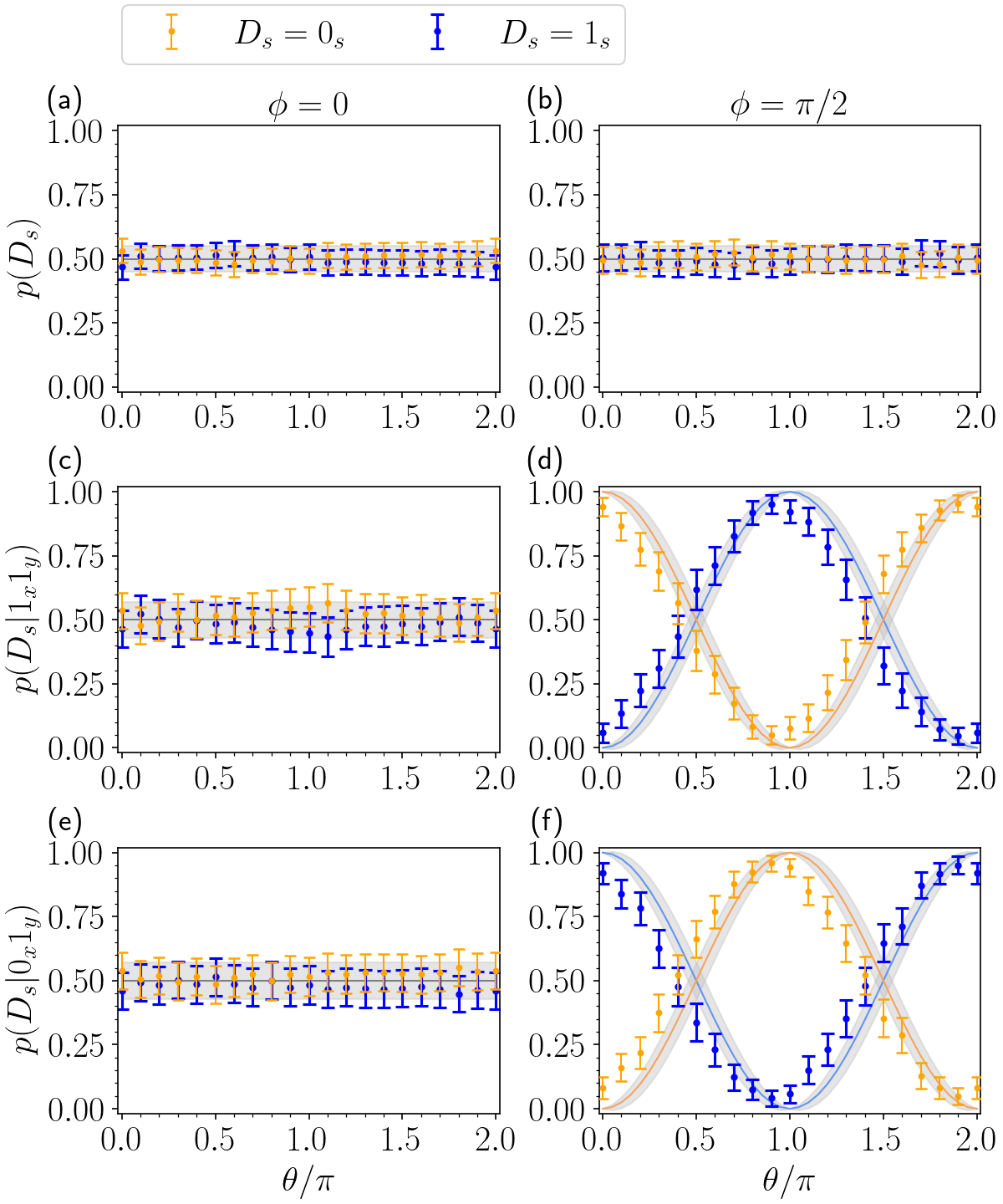}
\caption{Probabilities $P(D_s)$, $P(D_s|1_x1_y)$, and $P(D_s|0_x1_y)$ for $D_s=0_s$ and $D_s=1_s$ performed on \texttt{ibm\_kyiv} for the random choice between $\phi = 0$ and $\phi = \pi/2$ (left panel) and between $\phi = 0$ and $\phi = \pi/4$ (right panel) with $t_{\text{delay}} = 0$.}
\label{fig:ibm9045delay0}
\end{figure}
\begin{figure}
    \includegraphics[width=0.63\textwidth, angle = 90]{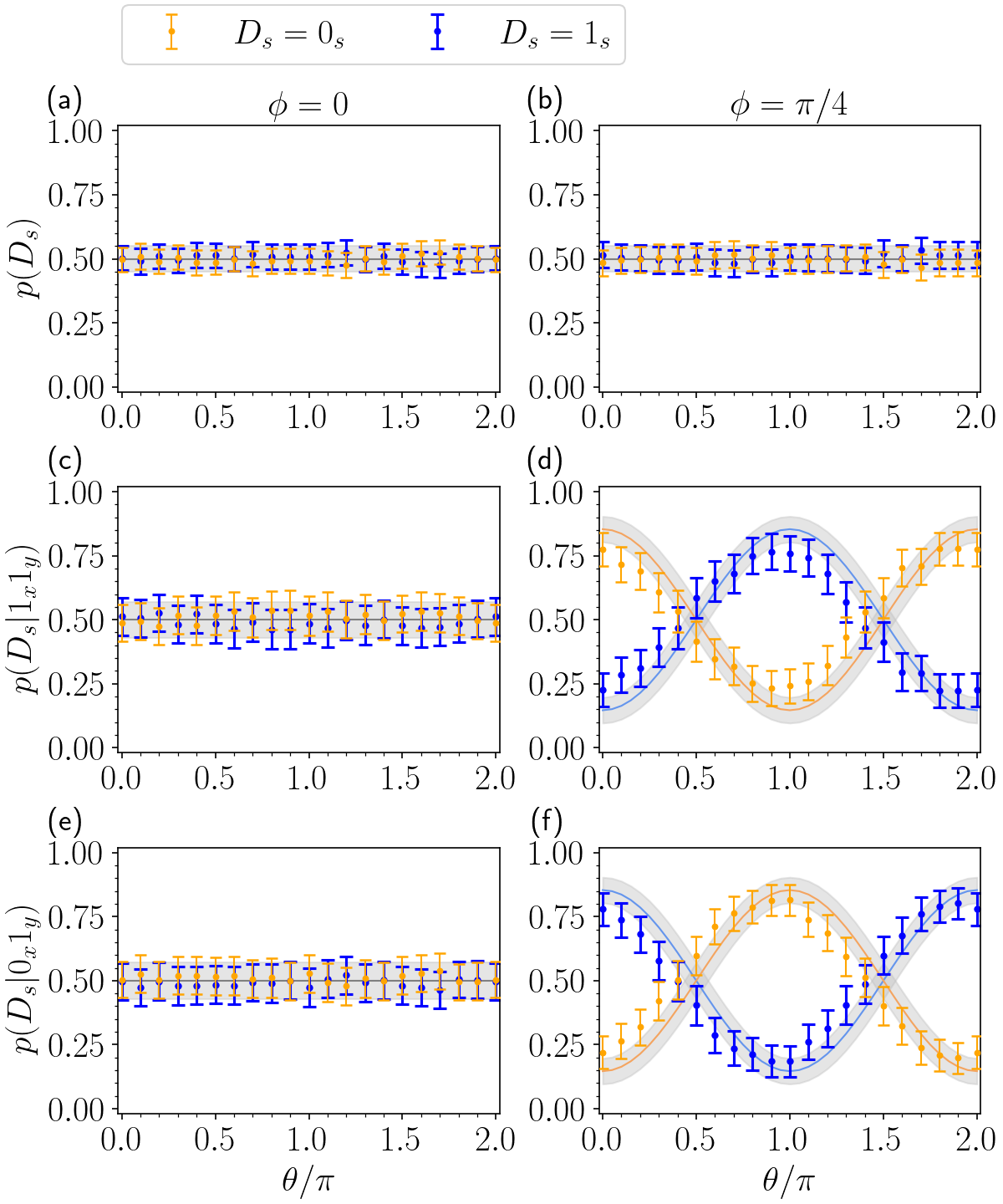}
    \vspace{0.6cm}
    \includegraphics[width=0.63\textwidth, angle = 90]{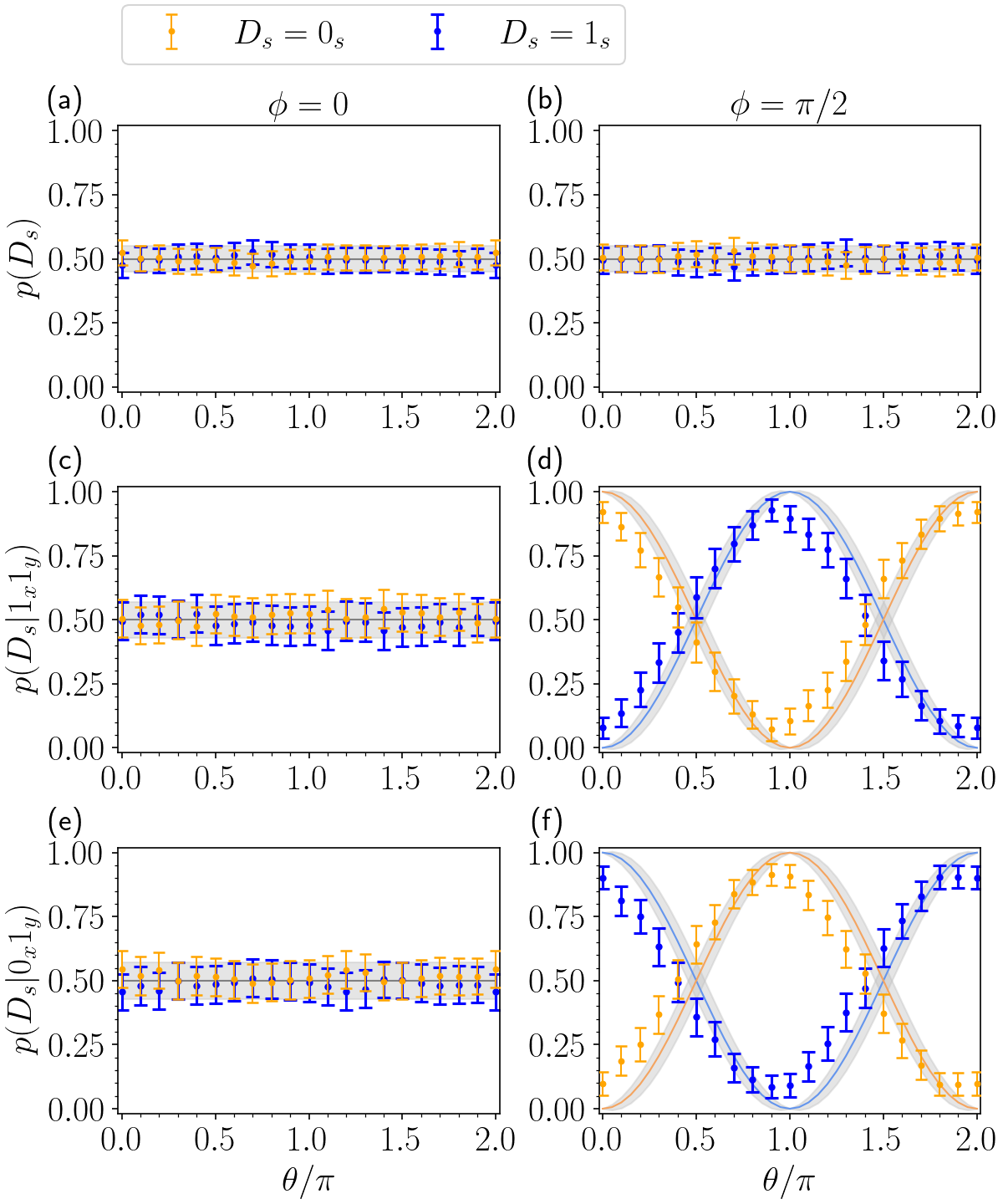}
\caption{Probabilities $P(D_s)$, $P(D_s|1_x1_y)$, and $P(D_s|0_x1_y)$ for $D_s=0_s$ and $D_s=1_s$ performed on \texttt{ibm\_kyiv} for the random choice between $\phi = 0$ and $\phi = \pi/2$ (left panel) and between $\phi = 0$ and $\phi = \pi/4$ (right panel) with $t_{\text{delay}} = 5{,}000\,dt \approx 1.11\,\mu\text{s}$.}
\label{fig:ibm9045delay5000}

\end{figure}

\begin{figure}
\centering
    \includegraphics[width=0.63\textwidth, angle =90]{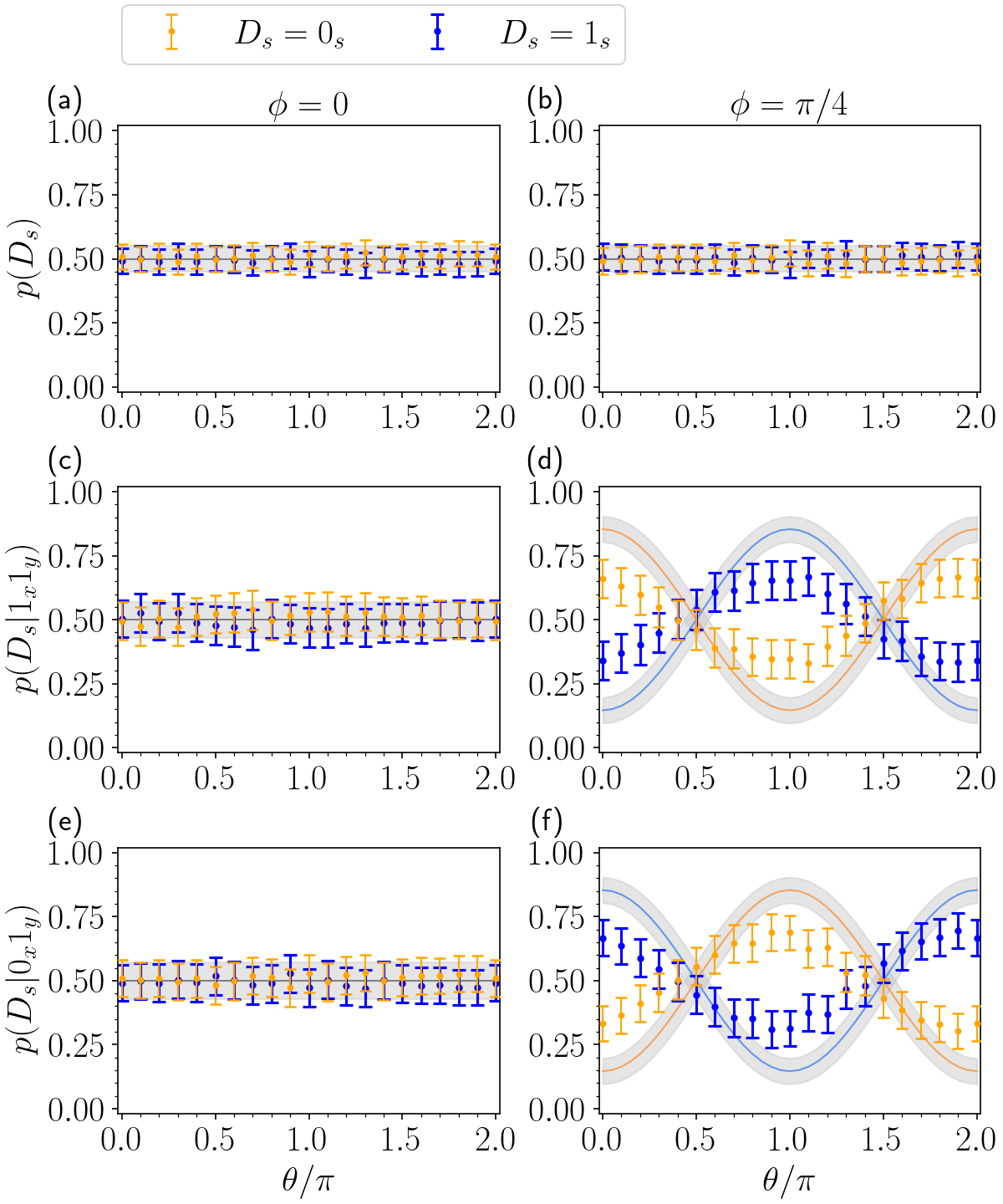}
    \vspace{0.6cm}
    \includegraphics[width=0.63\textwidth, angle = 90]{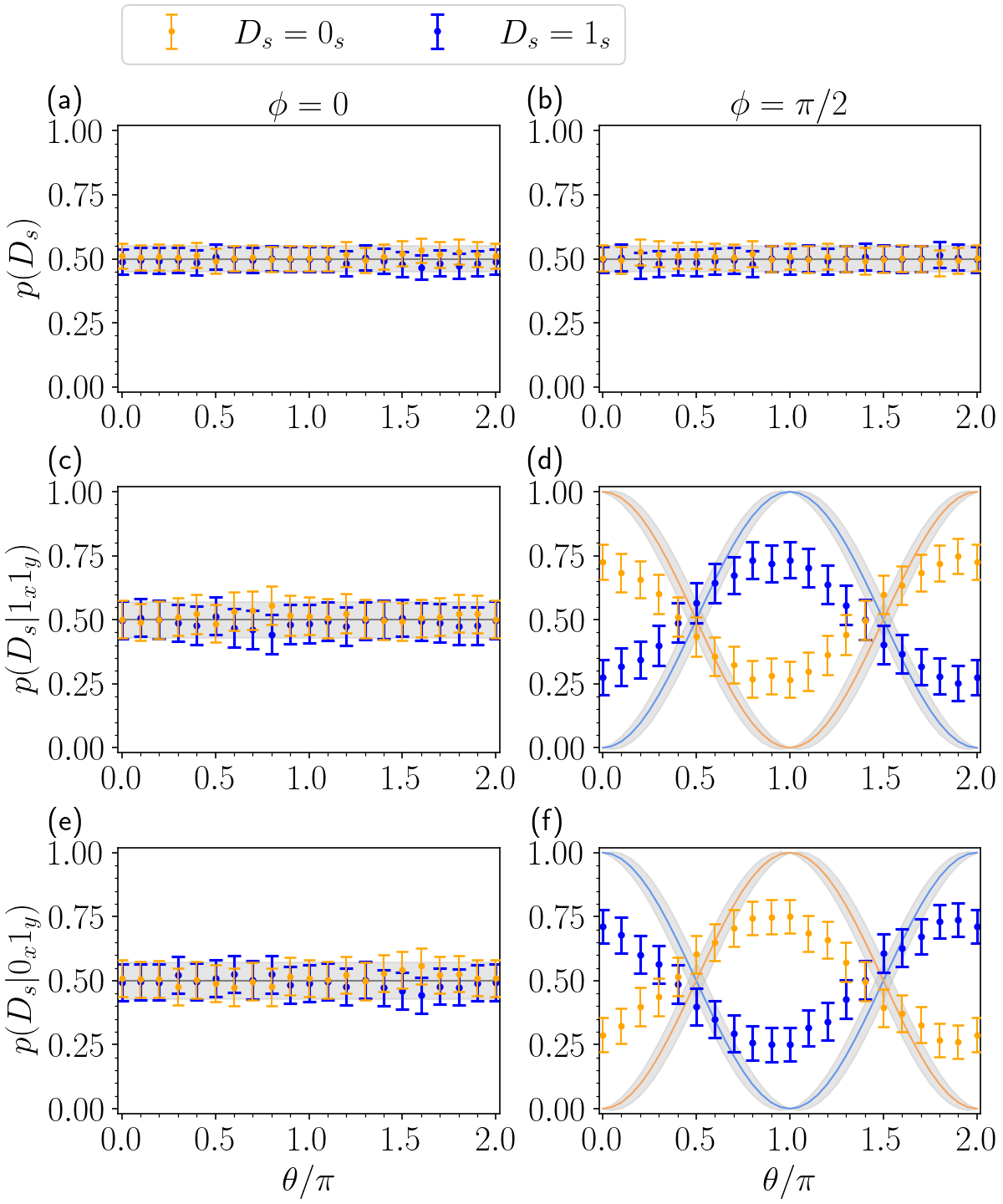}
\caption{Probabilities $P(D_s)$, $P(D_s|1_x1_y)$, and $P(D_s|0_x1_y)$ for $D_s=0_s$ and $D_s=1_s$ performed on \texttt{ibm\_kyiv} for the random choice between $\phi = 0$ and $\phi = \pi/2$ (left panel) and between $\phi = 0$ and $\phi = \pi/4$ (right panel) with $t_{\text{delay}} = 25{,}000\,dt \approx 5.56\,\mu\text{s}$.}
\label{fig:ibm9045delay25000}
\end{figure}
\begin{figure}
    \includegraphics[width=0.63\textwidth, angle = 90]{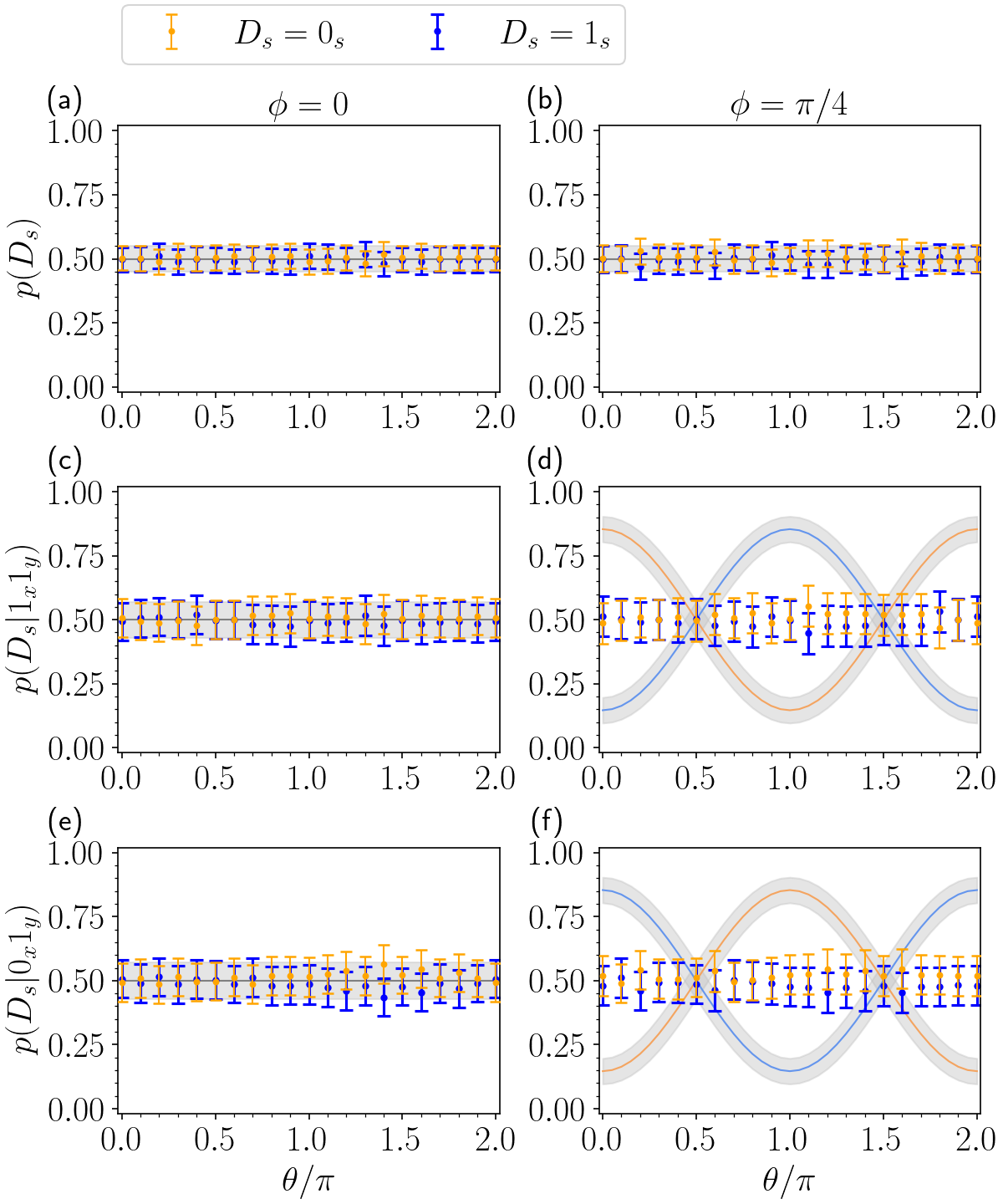}
    \vspace{0.6cm}
    \includegraphics[width=0.63\textwidth, angle = 90]{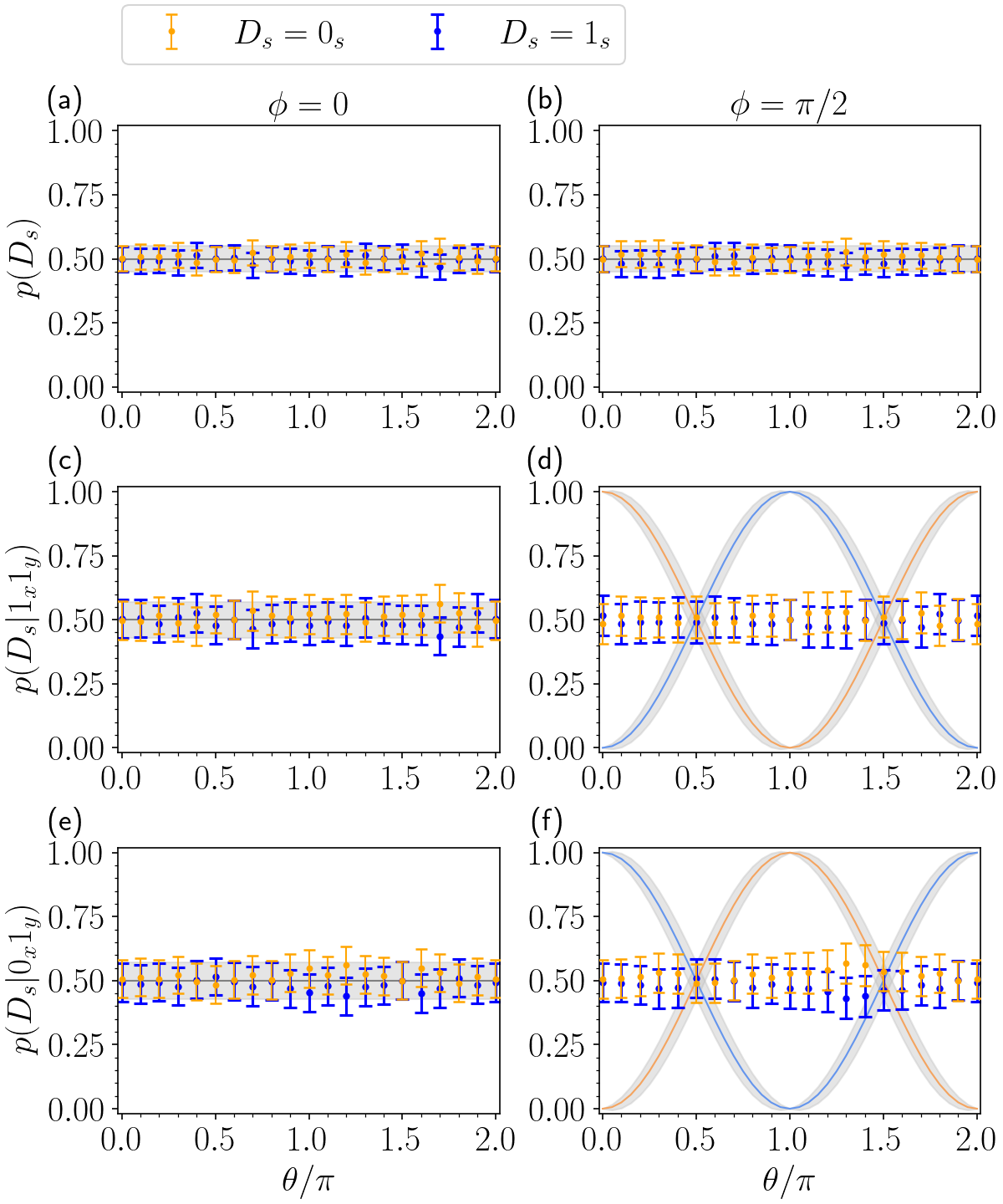}
\caption{Probabilities $P(D_s)$, $P(D_s|1_x1_y)$, and $P(D_s|0_x1_y)$ for $D_s=0_s$ and $D_s=1_s$ performed on \texttt{ibm\_kyiv} for the random choice between $\phi = 0$ and $\phi = \pi/2$ (left panel) and between $\phi = 0$ and $\phi = \pi/4$ (right panel) with $t_{\text{delay}} = 40{,}000\,dt \approx 8.89\,\mu\text{s}$.}
\label{fig:ibm9045delay40000}

\end{figure}

Each time for the same given value of $\theta$ in \figref{fig:transpiled}, we perform 5{,}000 shots to accumulate measurement outcomes. For each shot, the measurements of $D_s$, $D_x$ and $D_y$ are performed in the computational $(0/1)$ basis. Given a fixed $\phi$, we incrementally vary the value of $\theta$ from $0$ to $2\pi$ with the resolution $0.1\pi$. As the open configuration does not exhibit any two-path interference, we focus solely on the closed configuration. Theoretically, according to \eqref{p s} and \eqref{p xy}, the measurement outcomes are independent of the value of $\theta$ and $\phi$. On the other hand, the probabilities that $D_s$ yields 0 and 1 manifest the two-path interference in response to $\theta$ when one considers the events associated with the $1_x1_y$ or $0_x1_y$ subensemble separately. Various probabilities such as $p(0_s)$, $p(0_s|1_x1_y)$, etc.\ in the closed configurations are counted as the relative frequencies of occurrence of the corresponding outcomes from the repetitive shots.\footnote{As noted in the paragraph after \eqref{p xy}, the leaked outcomes associated with $0_x0_y$ and $1_x0_y$ are discarded.}

The experimental results are presented in \figref{fig:ibm9045delay0}--\figref{fig:ibm9045delay40000} for the cases of $t_\text{delay}=0$, $t_\text{delay}=5{,}000\,dt$, $t_\text{delay}=25{,}000\,dt$, and $t_\text{delay}=40{,}000\,dt$, respectively, where $dt\approx0.22\,\text{ns}$ is the system cycle time.
The results with $\phi=\pi/2$, which is supposed to yield full erasure, are shown in the left panels, while those with $\phi=\pi/4$, which is supposed to yield partial erasure, are shown in the right panels.
In each panel, the left column (a, c, e) presents the results of the events when $CR_y(\phi)$ is not invoked (i.e, $D_a$ yields 0), whereas the right column (b, d, f) presents the results of the events when $CR_y(\phi)$ is invoked (i.e, $D_a$ yields 1).
The top row (a, b) presents the probabilities $p(0_s)$ and $p(1_s)\equiv1-p(0_s)$; the middle row (c, d) and bottom row (e, f) present the interference patterns of the events within the confines of the $1_x1_y$ and $0_x1_y$ subensembles, respectively.

The solid lines in the graphs represent the theoretical predictions obtained from \eqref{p s}, \eqref{p 1x1y}, and \eqref{p 0x1y}. The statistical fluctuations in these theoretical values are quantified by the standard deviation $\sigma_\text{th}$, as given in \eqref{sigma th}. To enhance visibility, we magnify $\sigma_\text{th}$ by a factor of 5 and depict it as a thin band overlaid on the solid line.
The sampling uncertainty is characterized by $\sigma_{\bar{x}}$, as defined in \eqref{SEM}, and is also magnified by a factor of 5. It is represented as an error bar for each data point.\footnote{Both $\sigma_\text{th}$ and $\sigma_{\bar{x}}$ are inherently small across all our experiments due to the large number of shots, making them nearly invisible. To improve visibility, we amplify them by a factor of 5 in all figures throughout this paper for both the IBM Quantum and IonQ results.}

The experimental results for $t_{\text{delay}} = 0$ and $t_{\text{delay}} = 5{,}000\,dt \approx 1.11\,\mu\text{s}$ align well with theoretical predictions but show noticeable systematic deviations that exceed the sampling uncertainty $\sigma_{\bar{x}}$. Additionally, the results for $t_{\text{delay}} = 0$ and $t_{\text{delay}} = 5{,}000\,dt$ are very close to each other, yielding nearly identical systematic errors despite the difference in $t_{\text{delay}}$. This suggests that these errors are not purely stochastic but predominantly systematic and can be significantly reduced with more advanced calibration techniques.

In contrast, for $t_{\text{delay}} = 25{,}000\,dt \approx 5.56\,\mu\text{s}$, the interference pattern recovered by erasing the which-way information is significantly diminished. For $t_{\text{delay}} = 40{,}000\,dt \approx 8.89\,\mu\text{s}$, the interference pattern is entirely unrecoverable. These results demonstrate that as the delay time $t_\text{delay}$ increases, coherence degrades, resulting in a diminished ability to recover interference patterns. The time scale of $\sim 8.89\,\mu\text{s}$ is shorter than but comparable to the dephasing time ($T_2$) of single-qubit gates on \texttt{ibm\_kyiv} as shown in Table~\ref{tab:calibration}.

In \appref{app:IBM 4 options}, we consider the random selection of four rotation angles, $\{0, \phi_1, \phi_2, \phi_1 + \phi_2\}$, by including two ancilla qubits. This enables the random erasure of the which-way information in four distinct degrees, all achievable in a \emph{single} circuit.
The experimental results for $\phi_1 = \pi/6$ and $\phi_2 = \pi/3$, performed on \texttt{ibm\_kyiv} with $t_\text{delay} = 0$, $t_{\text{delay}} = 5{,}000\,dt$, $t_{\text{delay}} = 25{,}000\,dt$, and $t_{\text{delay}} = 40{,}000\,dt$ are presented in \figref{fig:ibm3060delay0}--\figref{fig:ibm3060delay40000}, respectively. As expected, the results within a subensemble corresponding to a specific rotation angle exhibit the two-path interference to varying degrees. However, due to the significant increase in gate number when transpiled, the 4-option results in \figref{fig:ibm3060delay0}--\figref{fig:ibm3060delay40000} suffer from significantly more systematic errors compared to the two-option results in \figref{fig:ibm9045delay0}--\figref{fig:ibm9045delay40000}.
However, the degradation of the recovered interference patterns with increasing delay time is less significant, indicating that coherence can be maintained for a longer duration compared to the two-option scenario. The comparison between the two-option and four-option cases further suggests that the errors are predominantly systematic.

\section{Experiments on IonQ}\label{sec:IonQ}
The experiments performed on \texttt{ibm\_kyiv} of IBM Quantum unavoidably invoke the SWAP operation, causing the $s$ qubit and the $x$ qubit to interchange their roles during computation, as shown in \figref{fig:transpiled}. This slightly weakens the resemblance to the Scully--Dr{\"u}hl-type quantum eraser.

To broaden the scope of experimental realizations of the Scully--Dr{\"u}hl-type quantum eraser, we also performed the experiment depicted in \figref{fig:quantum circuit} on the IonQ quantum processor \texttt{ionq\_harmony} \cite{IonQ}, provided by the Amazon Braket platform \cite{AWS}. The \texttt{ionq\_harmony} is an 11-qubit quantum processor in a trapped ion system, built on a chain of ${}^{171}\mathrm{Yb}^+$ ions in a microfabricated trap. Single-qubit and two-qubit gates are executed via a two-photon Raman transition by applying a pair of counter-propagating beams from a mode-locked pulsed laser. IonQ is fully connected, meaning that two-qubit gates can be performed between any pair of qubits. More details on the hardware technical specifications can be found in \cite{wright2019benchmarking}.

The all-to-all connectivity of IonQ eliminates the need for SWAP operations, allowing us to directly implement the circuit shown in \figref{fig:quantum circuit} alongside \figref{fig:random choice} without any swapping between the $s$, $x$, $y$, and $a$ qubits. The separation between neighboring ions is about 3--5\,$\mu\text{m}$ \cite{doi:10.1126/science.1231298, IonQ}, ensuring that the recorders of which-way information remain spatially separated by a distance on the order $\sim1\,\mu\text{m}$. However, since the state of all qubits is read out simultaneously by directing a resonant laser, we cannot perform the measurement of $D_s$ before the $CR_y(\phi)$ gate is invoked. Despite this limitation, the experiments on IonQ still demonstrate the Scully--Dr{\"u}hl-type quantum eraser, although the erasure is not achieved in a truly delayed-choice manner (but see Footnote~\ref{foot:delayed choice} for an alternative interpretation).

\begin{figure}
\centering
    \includegraphics[width=0.63\textwidth, angle=90]{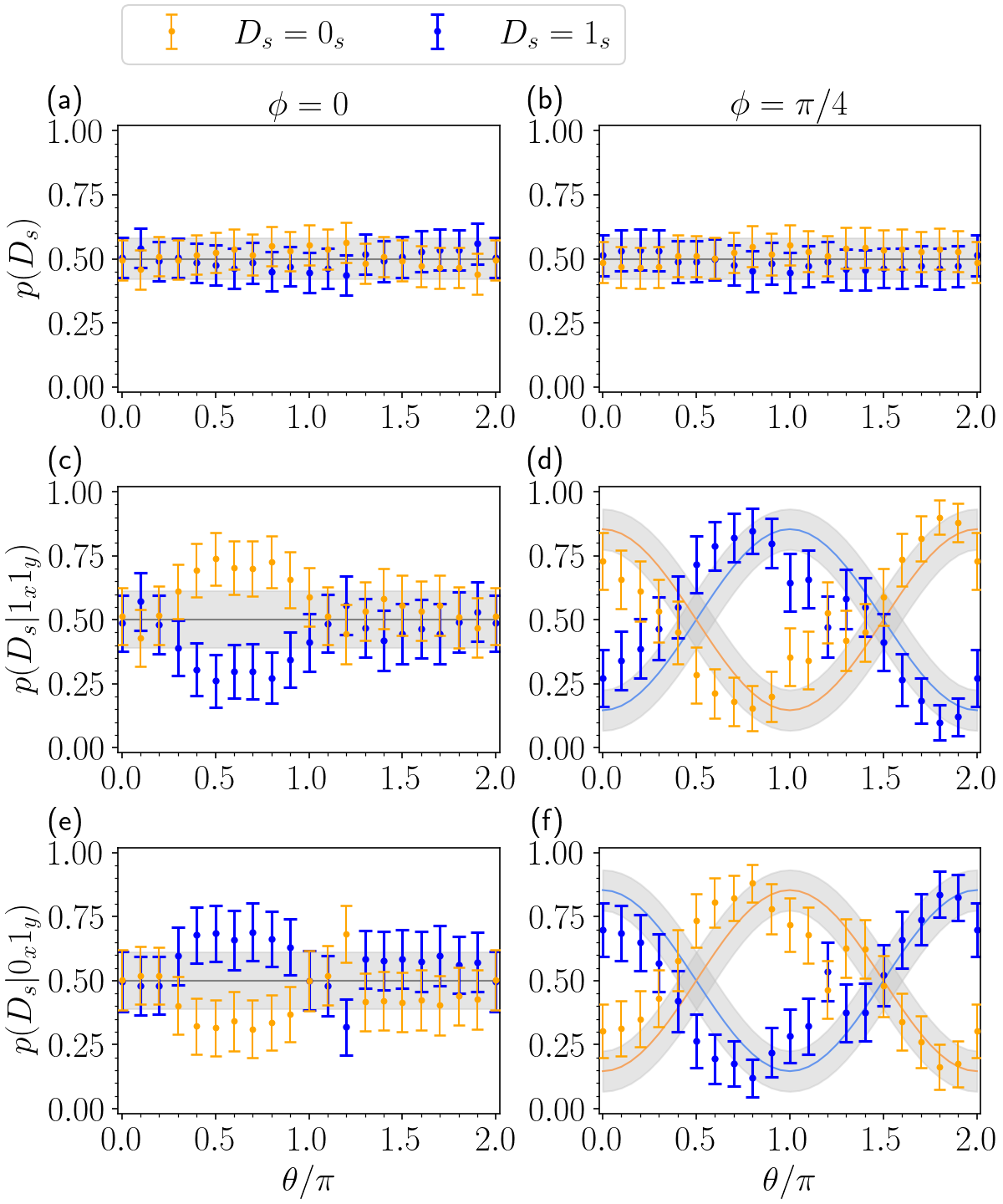}
    \vspace{0.6cm}
    \includegraphics[width=0.63\textwidth, angle=90]{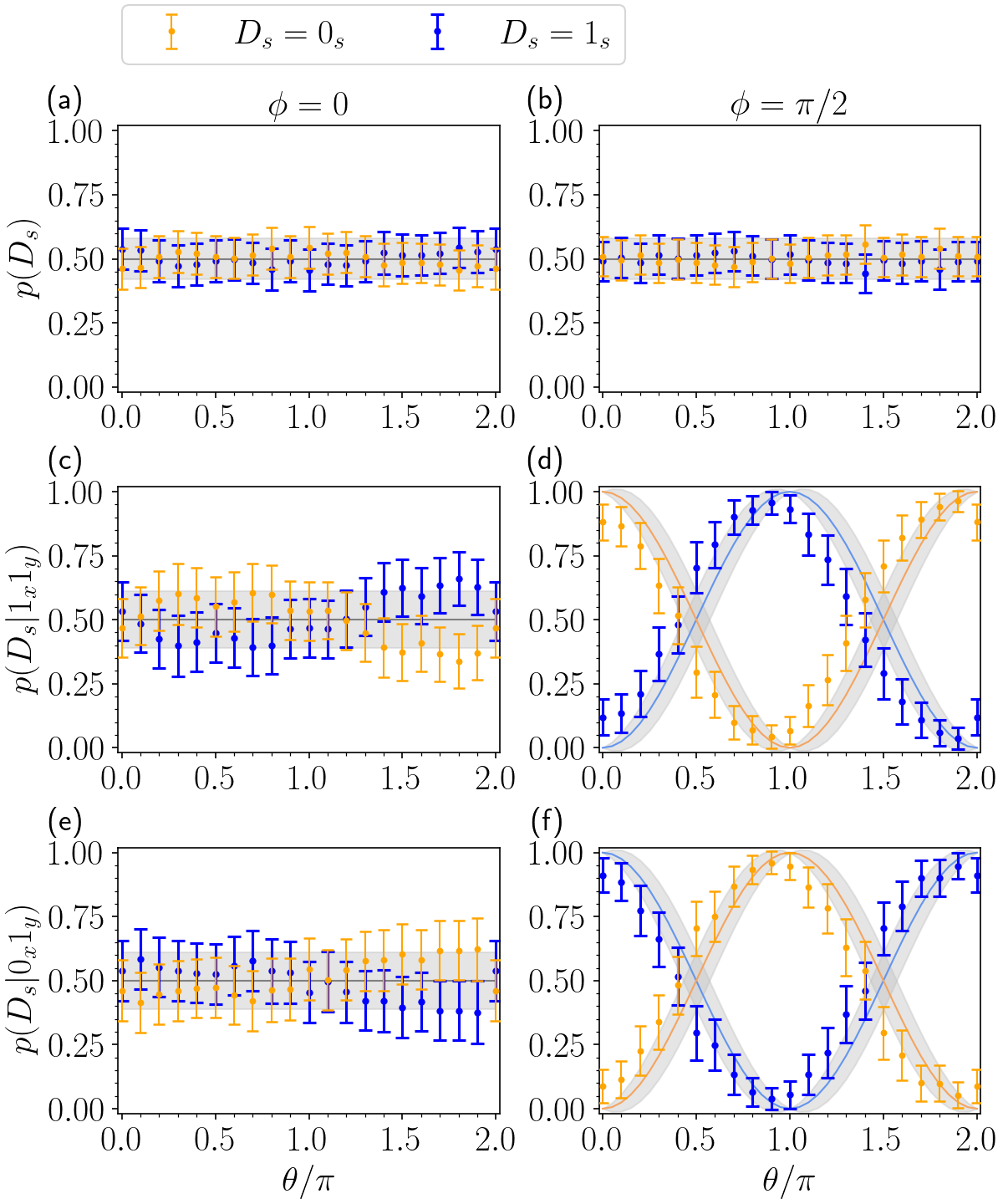}
\caption{Probabilities $P(D_s)$, $P(D_s|1_x1_y)$, and $P(D_s|0_x1_y)$ for $D_s=0_s$ and $D_s=1_s$ performed on \texttt{ionq\_harmony} for the random choice between $\phi =0$ and $\pi/2$ (left panel) and between $\phi =0$ and $\pi/4$ (right panel).}
\label{fig:ionq9045}
\end{figure}

The experimental results for $\phi = \pi/2$ and $\phi = \pi/4$ are shown in \figref{fig:ionq9045}, with 2{,}000 shots collected for each data point. Both sets of results align with theoretical predictions but exhibit considerable systematic errors. Compared to the baseline results without the random-choice mechanism, presented in \figref{fig:ionQ_interference}, these systematic errors are likely caused by the inclusion of the random-choice component, as depicted in \figref{fig:random choice}. When transpiled into primitive single-qubit gates and M{\o}lmer–S{\o}rensen (MS) gates (the only native two-qubit gates on the IonQ device), the random-choice component significantly increases circuit complexity, contributing to systematic errors that have yet to be properly calibrated.

In \appref{app:IonQ 4 options}, we also perform an experiment involving the random choice of four rotation angles, $\{0, \pi/6, \pi/3, \pi/2\}$, by including two ancilla qubits. The experimental results are presented in \figref{fig:ionq_3060}. The four-option results exhibit considerable systematic errors, but the errors are not necessarily more significant than those observed in the two-option results. This suggests that the errors encountered on \texttt{ionq\_harmony} are primarily systematic rather than purely stochastic and might be further reduced through more sophisticated calibration.

\section{Summary and discussion}\label{sec:summary}
We propose a quantum circuit, as shown in \figref{fig:quantum circuit}, which implements the Scully--Dr{\"u}hl-type delayed-choice quantum eraser in a genuine manner, ensuring that the two recorders of the which-way information make direct contact with the signal qubit while remaining spatially separated from each other.
This quantum circuit experiment is not only easier to implement than optical experiments but also facilitates arbitrary adjustment of the degree of erasure by simply tuning the rotation angle $\phi$ of the $R_y(\phi)$ gate.
Furthermore, we can achieve a true random choice in selecting different values of $\phi$ by utilizing the circuit component depicted in \figref{fig:random choice}.
We performed these experiments on the \texttt{ibm\_kyiv} processor of IBM Quantum and the \texttt{ionq\_harmony} processor of IonQ. The two qubits serving as recorders remain spatially separated by distances on the order of $\sim10^2\,\mu\text{m}$ and $\sim1\,\mu\text{m}$, respectively. Our experiments extend the experimental realizations of the genuine Scully--Dr{\"u}hl-type quantum eraser beyond optical experiments to those involving superconducting transmons and trapped ions, thereby broadening the scope of quantum erasure implementations.

In IBM Quantum processors, the measurement $D_s$ of the signal qubit can be performed midway, before the random choice is invoked, ensuring that the choice genuinely occurs in a delayed-choice manner. Furthermore, delay gates can be applied, as shown in \figref{fig:transpiled}, to further postpone the random choice, thereby amplifying the retrocausal effect. Since gate operations are executed sequentially in time, the system does not have any involvement of random choice until after the signal qubit has been measured. This approach eliminates any potential philosophical loopholes regarding retrocausality that might be present in other experimental setups.

The experiments conducted on \texttt{ibm\_kyiv} with varying delay times are shown in \figref{fig:ibm9045delay0}--\figref{fig:ibm9045delay40000}.
For $t_{\text{delay}} = 0$ and $t_{\text{delay}} = 5{,}000\,dt \approx 1.11\,\mu\text{s}$, the results are nearly identical and align well with theoretical predictions, displaying similar systematic errors despite the difference in $t_{\text{delay}}$. This indicates that the errors are predominantly systematic rather than stochastic.
For $t_{\text{delay}} = 25{,}000\,dt \approx 5.56\,\mu\text{s}$, the interference pattern significantly diminishes, and for $t_{\text{delay}} = 40{,}000\,dt \approx 8.89\,\mu\text{s}$, it becomes entirely unrecoverable.
These findings demonstrate that as $t_{\text{delay}}$ increases, coherence degrades, leading to a reduced ability to recover interference patterns. Remarkably, quantum erasure can be achieved with delay times up to approximately $1\,\mu\text{s}$ without noticeable decoherence. Achieving a similar delay in optical experiments would be highly challenging, as a delay of $1\,\mu\text{s}$ corresponds to a substantial distance of about $300\,\text{m}$.

To broaden the scope of experimental realizations, we also perform experiments on the IonQ processor \texttt{ionq\_harmony}, where the all-to-all connectivity eliminates the need for any SWAP operations that interchange the roles of different qubits midway. Unfortunately, in IonQ processors, all qubits are measured simultaneously at the end of the entire computation, rendering it impossible to perform the measurement of $D_s$ before the random choice. Nevertheless, one can still assert that the erasure is achieved in a delayed-choice manner on the grounds that the random choice is not finalized until the state of the ancilla qubit is known, which occurs when the measurement $D_a$ is performed (recall Footnote~\ref{foot:delayed choice}). The experiments conducted on \texttt{ionq\_harmony} are presented in \figref{fig:ionQ_interference} for the case without the random choice, and in \figref{fig:ionq9045} for the case with the random choice. The experimental results in \figref{fig:ionQ_interference} agree closely with the theoretical predictions. By comparison, the experimental results in \figref{fig:ionq9045} are in good agreement with the theoretical predictions but exhibit some noticeable systematic errors.

The random choice circuit can be extended to include multiple options within a single circuit, as illustrated in \figref{fig:random choice of 4 options}. In Appendix~\ref{app:4 options}, we consider the random choice of four options $\phi \in \{0, \pi/6, \pi/3, \pi/2\}$. The experiments conducted on \texttt{ibm\_kyiv} are presented in \figref{fig:ibm3060delay0}--\figref{fig:ibm3060delay40000}, and those conducted on \texttt{ionq\_harmony} are presented in \figref{fig:ionq_3060}.

Compared to their two-option counterparts, the four-option results shown in \figref{fig:ibm3060delay0}--\figref{fig:ibm3060delay40000} exhibit more systematic errors but demonstrate milder decoherence as the delay time increases.
On the other hand, the four-option results in \figref{fig:ionq_3060} also display considerable systematic errors, but not necessarily more significant than the two-option counterparts.
Except for the no-random-choice results, as shown in \figref{fig:ionQ_interference}, which closely agree with the theory, all other two-option and four-option experiments on both \texttt{ibm\_kyiv} and \texttt{ionq\_harmony} display noticeable characteristic deviations from the theoretical predictions. This suggests that the errors are not purely stochastic but mainly systematic: gate errors do not occur independently across different gates but are somehow correlated through two-qubit gates, leading to unwanted correlations when more qubits are involved and thus resulting in systematic errors of varying degrees. Since these errors are mainly systematic, they can be greatly mitigated through more sophisticated calibration based on a more meticulous analysis of the error correlation.

If the systematic errors can be further calibrated, the interference pattern recovered by erasing the which-way information to any desired degree will closely match the predicted level. As the fidelity of quantum processors from IBM Quantum, IonQ, and other architectures continues to improve, quantum circuits will provide an effective platform for conducting various quantum experiments. This serves as a reliable alternative to optical experiments, which often suffer from unwanted sources of decoherence.
In addition to the entanglement quantum eraser experiments demonstrated in \cite{chiou2024complementarity} and the Scully--Dr{\"u}hl-type quantum eraser experiments demonstrated in this paper, advances in quantum circuit technology will open new and promising avenues for exploring more quantum effects, potentially transforming the landscape of quantum research.

\begin{acknowledgments}
The authors would like to thank Jie-Hong R.\ Jiang and the anonymous referees for their valuable comments and suggestions, which greatly improved the quality of this manuscript. H.C.H.\ also thanks Yen-Hsiang Lin for insightful discussions.
This work was supported in part by the National Science and Technology Council of Taiwan through Grants 111-2112-M-110-013, 112-2119-M-007-008, 112-2119-M-002-017, 113-2119-M-002-024, and 113-2119-M-007-013, as well as by the NTU Center for Data Intelligence: Technologies, Applications, and Systems under Grant NTU-113L900903. We gratefully acknowledge the IBM Q Hub at NTU for providing access to IBM Quantum resources and the Center for Quantum Frontiers of Research and Technology at NCKU for providing access to IonQ.
\end{acknowledgments}

\appendix

\section{Experiments without random choice}\label{app:no choice}
We also conducted experiments without the random choice. This approach avoids additional errors introduced by the ancillary qubit circuit, providing a baseline for comparison. The value of $\phi$ was preset to $0$, $\pi/4$, and $\pi/2$, respectively.

\begin{figure}
\centering
    \includegraphics[width=1.25\textwidth, angle =90 ]{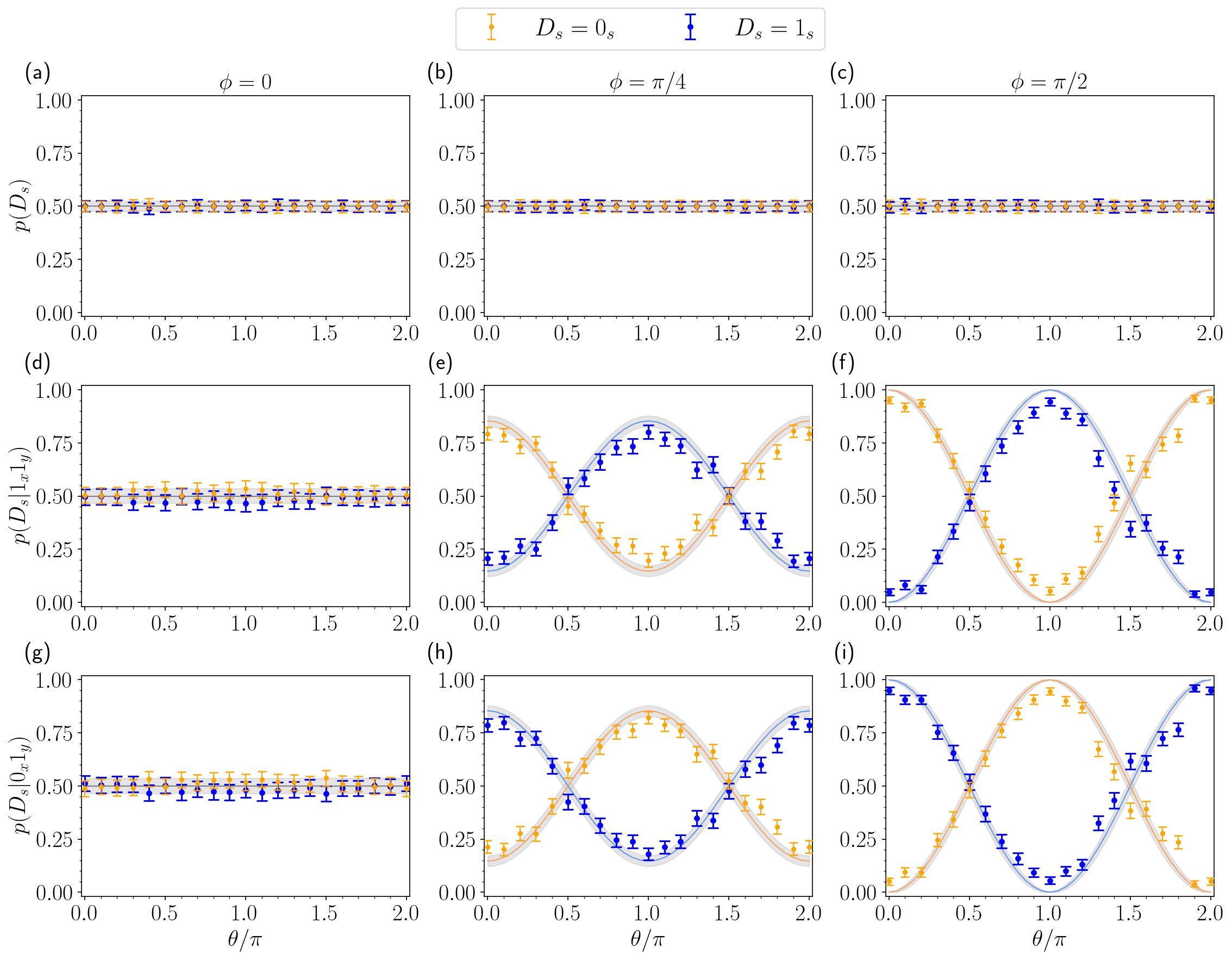}
\caption{Probabilities $P(D_s)$, $P(D_s|1_x1_y)$, and $P(D_s|0_x1_y)$ for $D_s=0_s$ and $D_s=1_s$ performed on \texttt{ibm\_kyiv} for the cases of $\phi=0$ (left column), $\phi=\pi/4$ (middle column), and $\phi=2\pi$ (right column) without the random choice.}
\label{fig:ibm_interference}
\end{figure}

\begin{figure}
    \includegraphics[width=1.25\textwidth, angle =90 ]{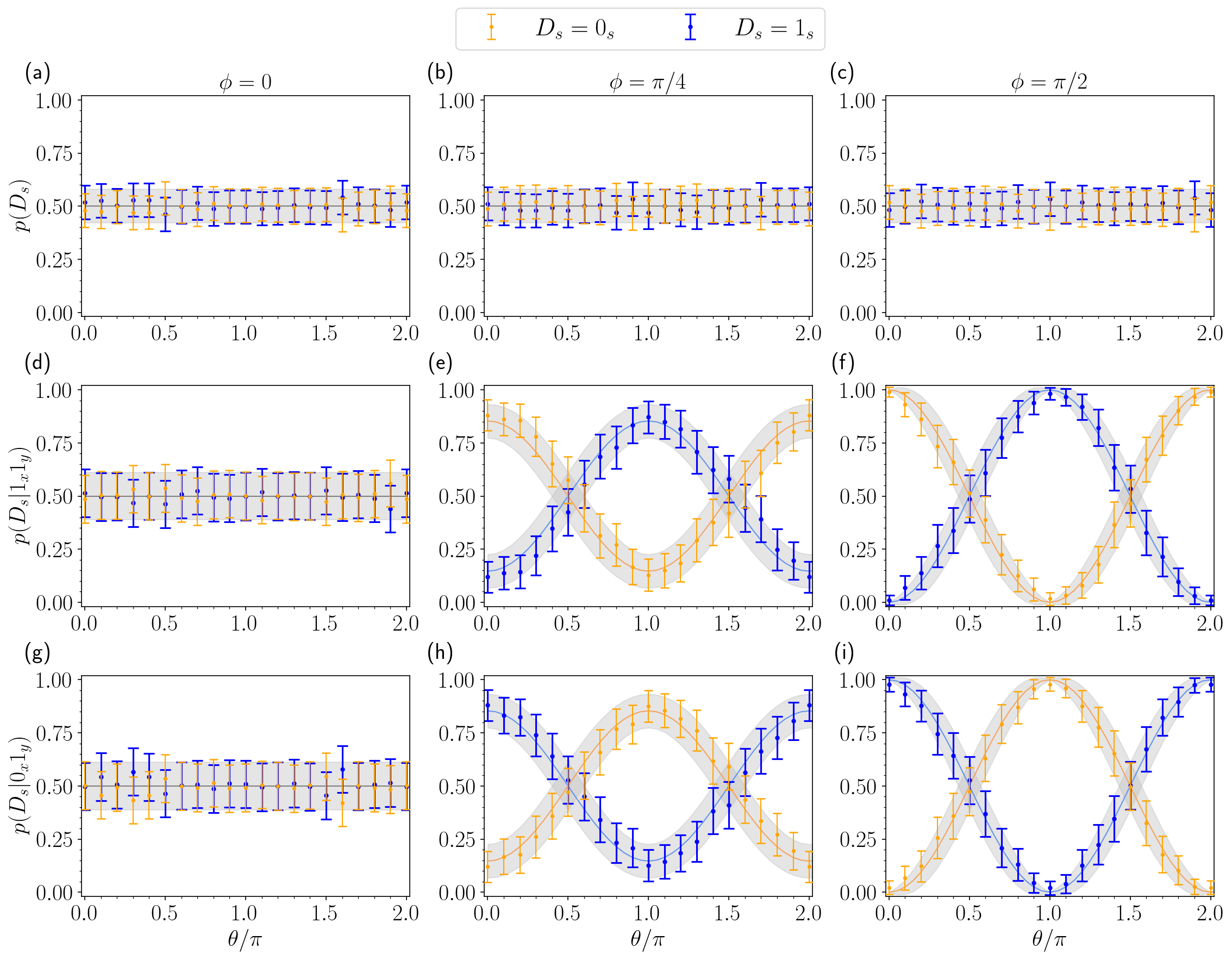}
\caption{Probabilities $P(D_s)$, $P(D_s|1_x1_y)$, and $P(D_s|0_x1_y)$ for $D_s=0_s$ and $D_s=1_s$ performed on \texttt{ionq\_harmony} for the cases of $\phi=0$ (left column), $\phi=\pi/4$ (middle column), and $\phi=2\pi$ (right column) without the random choice.}
\label{fig:ionQ_interference}
\end{figure}

The experimental results on \texttt{ibm\_kyiv} are shown in \figref{fig:ibm_interference}, with each data point obtained from 10{,}000 shots. Similarly, the results from \texttt{ionq\_harmony} are displayed in \figref{fig:ionQ_interference}, with each data point is obtained from 1{,}000 shots. In both cases, the results show good agreement with theoretical predictions.

\section{Experiments with four-option random choice}\label{app:4 options}
In the main body of this paper, we employ the circuit component illustrated in \figref{fig:random choice} to automatically perform the random choice between two options, $\{0, \phi\}$, for the rotation angle with equal probability. With a minor adjustment, this approach can be extended to facilitate random selection among multiple options within a \emph{single} circuit.
Specifically, by including two ancilla qubits, $a_1$ and $a_2$, we can design a circuit that randomly selects, with equal probability, one of four rotation angles $\{0, \phi_1, \phi_2, \phi_1 + \phi_2\}$, as shown in \figref{fig:random choice of 4 options}. The chosen rotation angle can be determined by the combined outcomes of the measurements $D_{a_1}$ and $D_{a_2}$.
In the following, we present the experimental results with we $\phi_1 = \pi/6$ and $\phi_2 = \pi/3$, resulting in four options $\{0, \pi/6, \pi/3, \pi/2\}$.

\begin{figure}
\begin{quantikz}
& \lstick{$\cdots$}  & \gate{R_y(\phi_1)} & \gate{R_y(\phi_2)} & \meter{$D_x$} \\
\lstick{$\ket{0}_{a_1}$} & \gate{H} & \ctrl{-1} & \qw & \meter{$D_{a_1}$} \\
\lstick{$\ket{0}_{a_2}$} & \gate{H} & \qw & \ctrl{-2} & \meter{$D_{a_2}$}
\end{quantikz}
\caption{The circuit component for the random choice of applying four different rotation angles.}
\label{fig:random choice of 4 options}
\end{figure}

\subsection{On IBM Quantum}\label{app:IBM 4 options}
Since the processor \texttt{ibm\_kyiv} is not fully connected, the SWAP operation is necessary to implement the quantum eraser, as shown in \figref{fig:transpiled} for the two-option random choice. To extend this to a four-option random choice, we replace the sub-circuit of \figref{fig:layout} that includes the entire ancilla-qubit wire, the $CR_y(\phi)$ gate, and the measurement $D_x$ with the circuit component depicted in \figref{fig:4 choices}.

\begin{figure}
\begin{quantikz}
                           & \lstick{$\cdots$}  & \gate{R_y(\phi_2)}&\qw        &\gate{R_y(\phi_1)} & \meter{$D_x$}\\
\lstick{$\ket{0}_{a_2}$}   & \gate{H}           & \ctrl{-1}         &\swap{1}   &\ctrl{-1}          &\meter{$D_{a_1}$}\\
\lstick{$\ket{0}_{a_1}$}   & \gate{H}           & \qw               &\swap{-1}  &\qw                &\meter{$D_{a_2}$}
\end{quantikz}
\caption{The circuit component for the random choice of applying four different rotation angles on \texttt{ibm\_kyiv}.}
\label{fig:4 choices}
\end{figure}

\begin{figure}
\centering
    \includegraphics[width=1.35\textwidth, angle =90 ]{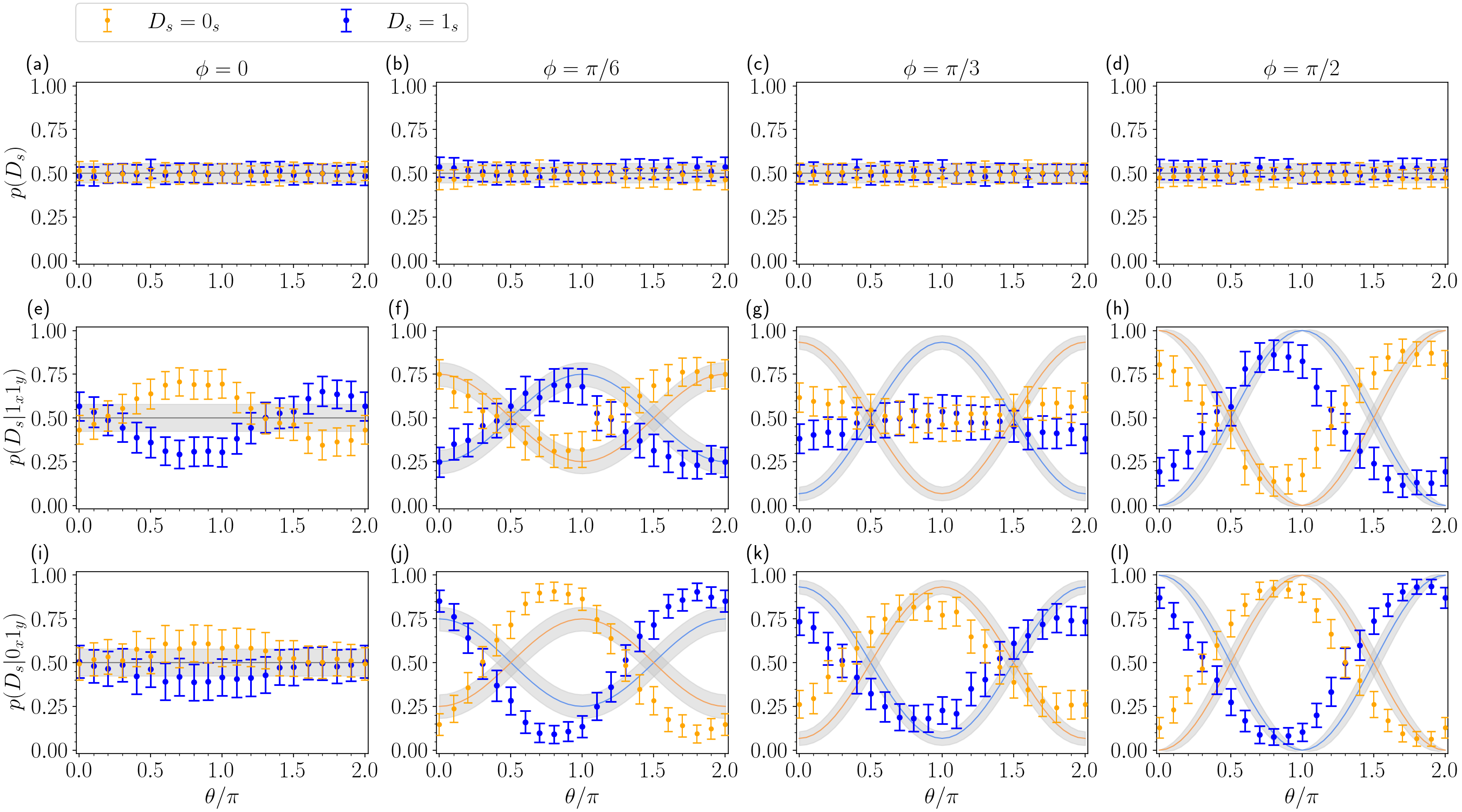}
\caption{Probabilities $P(D_s)$, $P(D_s|1_x1_y)$, and $P(D_s|0_x1_y)$ for $D_s=0_s$ and $D_s=1_s$ performed on \texttt{ibm\_kyiv} for the random choice between $\phi =0$, $\pi/6$, $\pi/3$, and $\pi/2$  with $t_\text{delay} = 0$.}
\label{fig:ibm3060delay0}
\end{figure}
\begin{figure}
    \includegraphics[width=1.35\textwidth, angle =90 ]{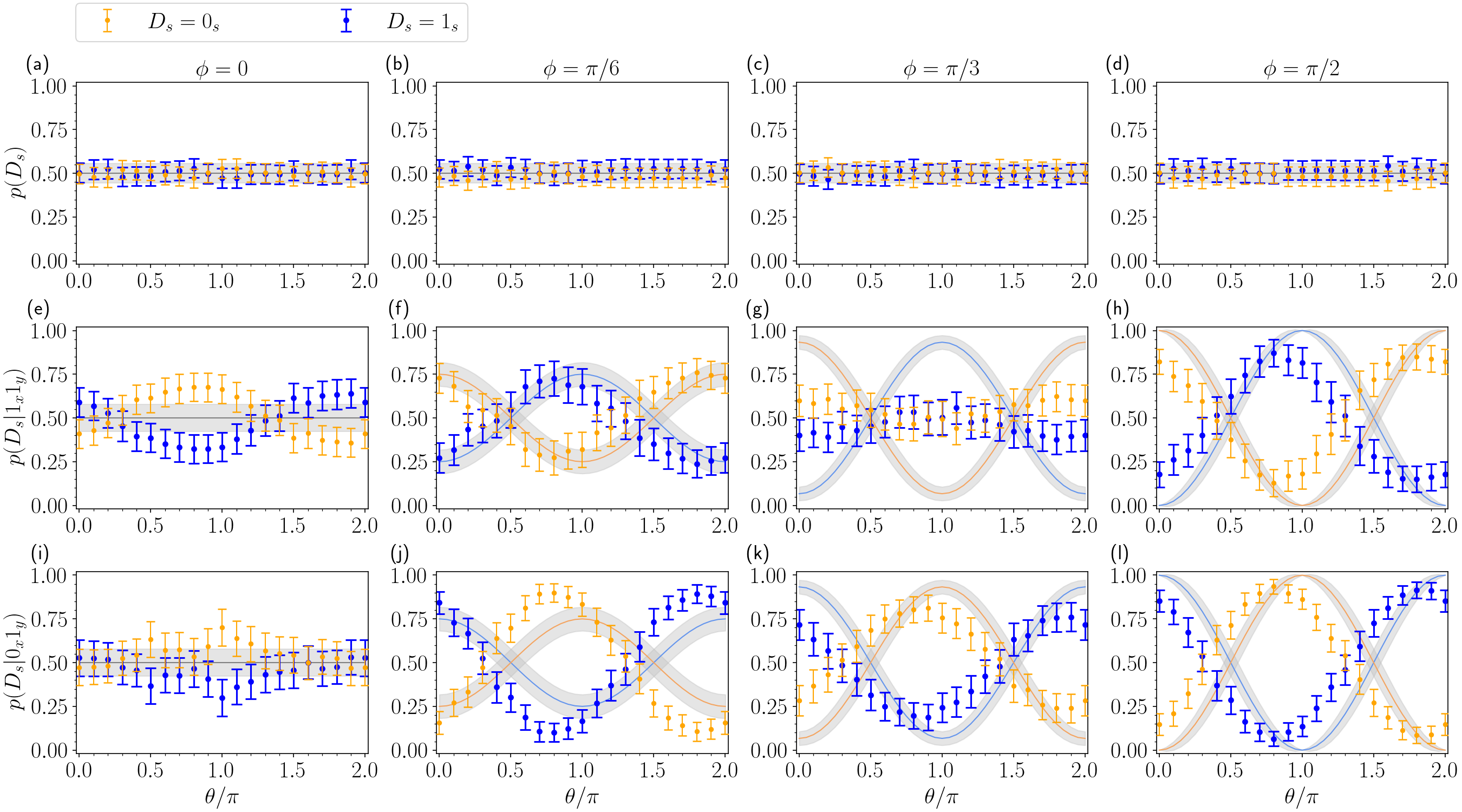}
\caption{Probabilities $P(D_s)$, $P(D_s|1_x1_y)$, and $P(D_s|0_x1_y)$ for $D_s=0_s$ and $D_s=1_s$ performed on \texttt{ibm\_kyiv} for the random choice between $\phi =0$, $\pi/6$, $\pi/3$, and $\pi/2$  with $t_{\text{delay}} = 5{,}000\,dt \approx 1.11\,\mu\text{s}$.}
\label{fig:ibm3060delay5000}

\end{figure}

\begin{figure}
\centering
    \includegraphics[width=1.35\textwidth, angle = 90]{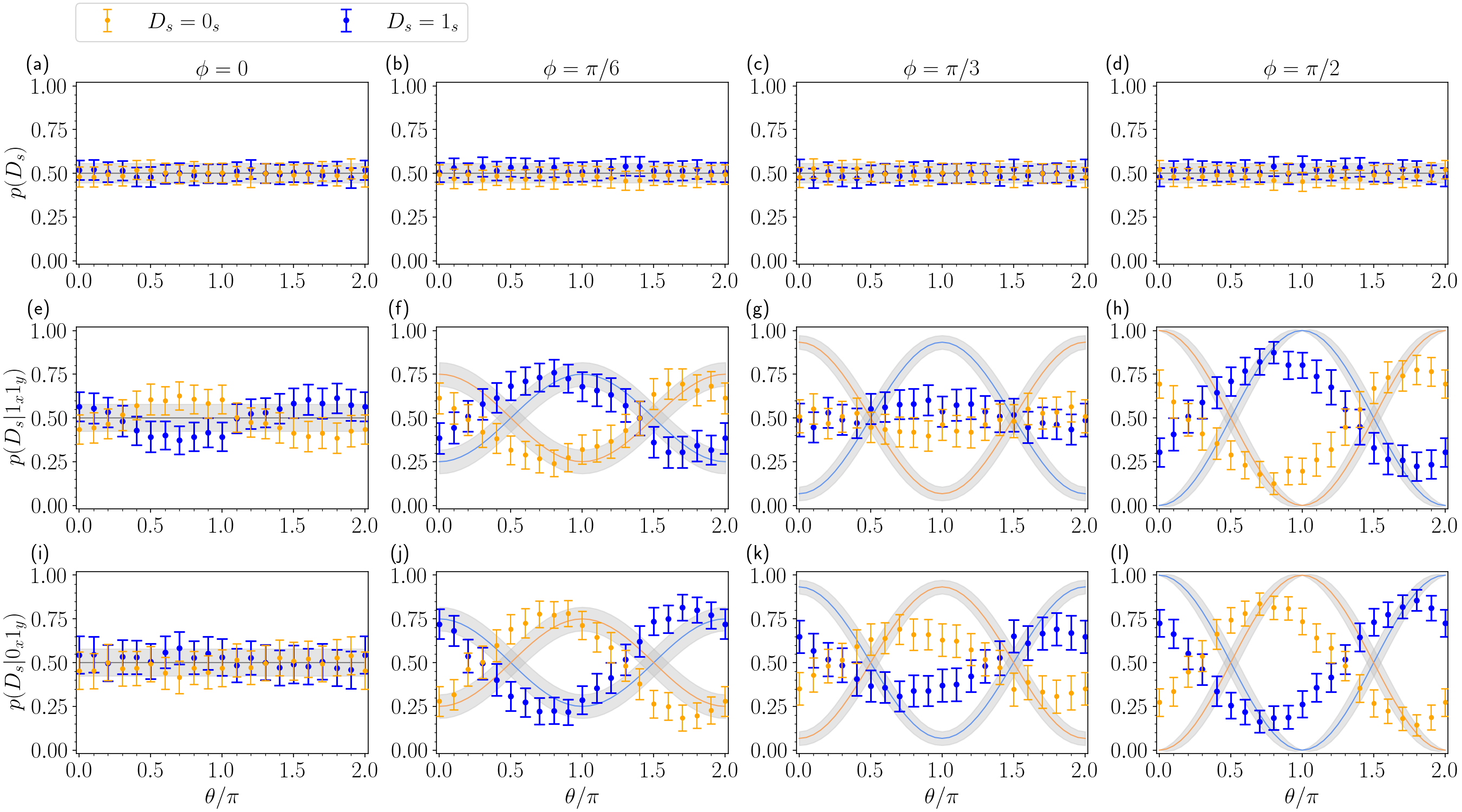}
\caption{Probabilities $P(D_s)$, $P(D_s|1_x1_y)$, and $P(D_s|0_x1_y)$ for $D_s=0_s$ and $D_s=1_s$ performed on \texttt{ibm\_kyiv} for the random choice between $\phi =0$, $\pi/6$, $\pi/3$, and $\pi/2$  with $t_{\text{delay}} = 25{,}000\,dt \approx 5.56\,\mu\text{s}$.}
\label{fig:ibm3060delay25000}
\end{figure}
\begin{figure}
    \includegraphics[width=1.35\textwidth, angle = 90]{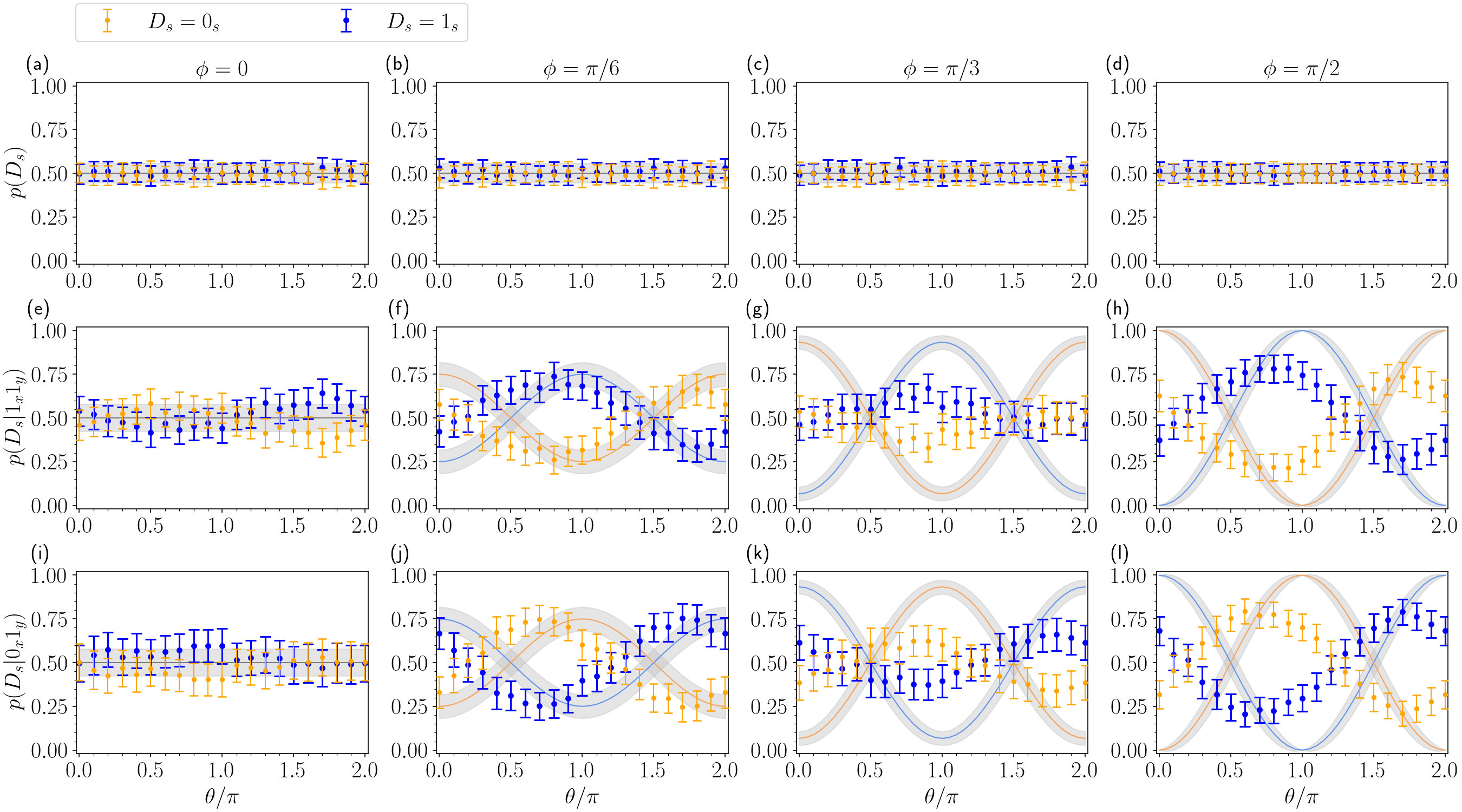}
\caption{Probabilities $P(D_s)$, $P(D_s|1_x1_y)$, and $P(D_s|0_x1_y)$ for $D_s=0_s$ and $D_s=1_s$ performed on \texttt{ibm\_kyiv} for the random choice between $\phi =0$, $\pi/6$, $\pi/3$, and $\pi/2$  with $t_{\text{delay}} = 40{,}000\,dt \approx 8.89\,\mu\text{s}$.}
\label{fig:ibm3060delay40000}

\end{figure}

For the experiments on \texttt{ibm\_kyiv}, we select qubits 92, 101, 102, 103, and 104 from \figref{fig:layout}, with the initial qubit mapping $(s,x,y,a_1,a_2) \mapsto (102,103,92,104,101)$,\footnote{Refer to \figref{fig:low level 4options delay5000} for the transpiled low-level circuit in terms of primitive gates and the exact gate execution schedule.} and perform 8{,}192 shots to accumulate measurement outcomes for each given value of $\theta$.
The experimental results are presented in \figref{fig:ibm3060delay0}--\figref{fig:ibm3060delay40000} for $t_\text{delay} = 0$, $t_{\text{delay}} = 5{,}000\,dt$, $t_{\text{delay}} = 25{,}000\,dt$, and $t_{\text{delay}} = 40{,}000\,dt$, respectively.
These results demonstrate that the two-path interference is recovered to varying degrees, corresponding to the different extents of which-way information erasure, all achieved within a single circuit.

Compared to the two-option results shown in \figref{fig:ibm9045delay0}--\figref{fig:ibm9045delay40000}, the four-option results exhibit noticeably more systematic errors. This is anticipated, as the circuit implementing four random choices requires considerably more gates in the transpiled low-level circuit compared to the two-option circuit.\footnote{This can be seen by comparing \figref{fig:low level 2options delay5000} with \figref{fig:low level 4options delay5000}.}
However, the recovered interference patterns are not significantly diminished even for $t_{\text{delay}} = 40{,}000\,dt$, indicating that coherence can be maintained for a longer duration compared to the two-option scenario.

\subsection{On IonQ}\label{app:IonQ 4 options}

\begin{figure}
\centering
    \includegraphics[width=1.35\textwidth, angle =90]{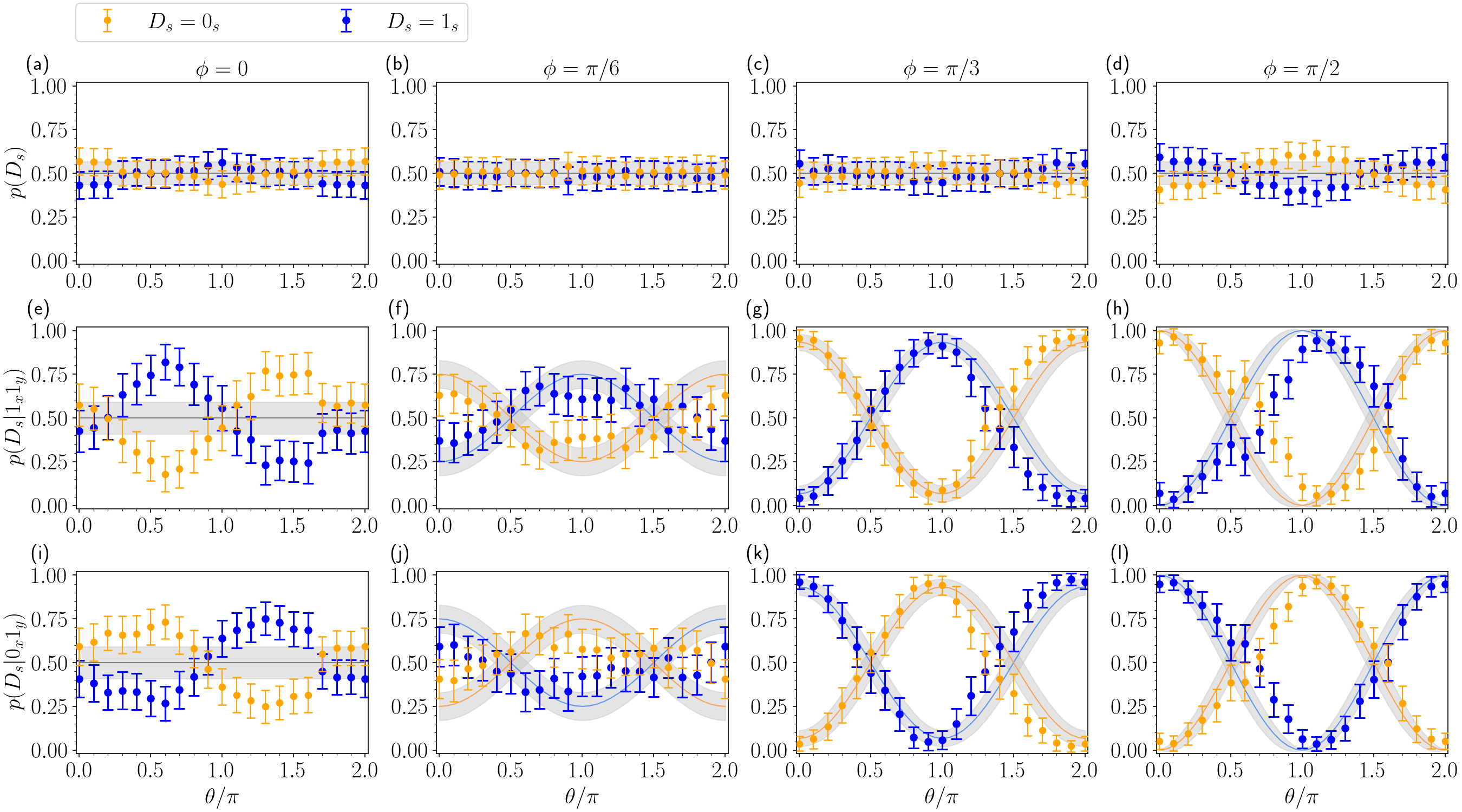}
\caption{Probabilities $P(D_s)$, $P(D_s|1_x1_y)$, and $P(D_s|0_x1_y)$ for $D_s=0_s$ and $D_s=1_s$ performed on \texttt{ionq\_harmony} for the random choice between $\phi =0$, $\pi/6$, $\pi/3$, and $\pi/2$.}
\label{fig:ionq_3060}
\end{figure}

The experimental results with four-option random choice conducted on \texttt{ionq\_harmony}, performing 6{,}000 shots for each given value of $\theta$, are presented in \figref{fig:ionq_3060}.
These results once again demonstrate that the random selection among multiple options within a single circuit can be achieved to erase the which-way information to multiple varying degrees.

When comparing the four-option results in \figref{fig:ionq_3060} to the two-option cases in \figref{fig:ionq9045}, the four-option results also exhibit considerable systematic errors, though not necessarily more pronounced than those in the two-option results. Notably, the results in the third column of \figref{fig:ionq_3060}, corresponding to $\phi = \pi/3$, show much smaller errors compared to the two-option case. This strongly suggests that, in experiments conducted on \texttt{ionq\_harmony}, the errors encountered are predominantly systematic rather than purely stochastic, highlighting the need for more sophisticated calibration.

\section{Technical details}\label{app:technical details}

\subsection{System calibration data}\label{app:calibration}
The system calibration data for \texttt{ibm\_kyiv} and \texttt{ionq\_harmony} are summarized in Table~\ref{tab:calibration}, including the relaxation time ($T_1$) and dephasing time ($T_2$) of single-qubit gates, the single-qubit gate error rate, and the two-qubit gate error rate. For \texttt{ibm\_kyiv}, we present the range of values for the physical qubits indexed from 0 to 4. For \texttt{ionq\_harmony}, we present the average values over all the qubits.

\begin{table}
\begin{tabular}{ |c|c|c|c|c| }
 \hline
 device &  $T_1$ & $T_2$ & 1-qubit gate error& 2-qubit gate error\\ \hline
 \texttt{ibm\_kyiv} & 108.95--293\,$\mu$s & 33.33--406.43\,$\mu$s& $1.17\times 10^{-4}$--$8.59 \times 10^{-4}$&  $7.32\times 10^{-3}$--$1.37\times 10^{-2}$\\ \hline
 \texttt{ionq\_harmony} & 10--100\,s& $\sim$1\,s & $4\times 10^{-3}$ &$ 2.7\times 10^{-2}$ \\
 \hline
\end{tabular}
\caption{Calibration data of \texttt{ibm\_kyiv} and \texttt{ionq\_harmony}.}
\label{tab:calibration}
\end{table}

\begin{figure}
\centering
\includegraphics[width=\textwidth]{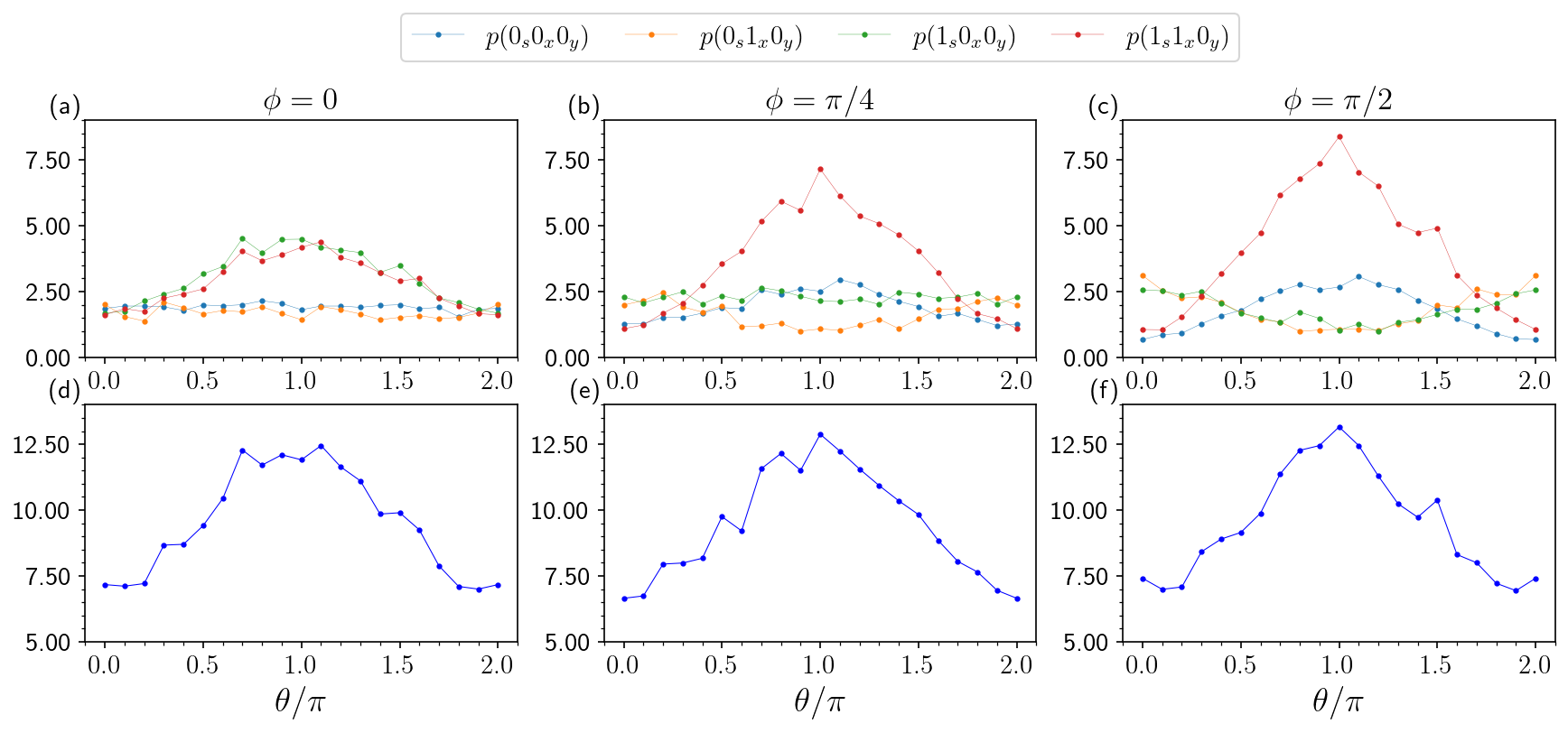}
\caption{Leakage probabilities (\%) of the experiment presented in \figref{fig:ibm_interference} on \texttt{ibm\_kyiv} without the random choice. The upper row displays individual probabilities for $p(0_s0_x0_y)$, $p(0_s1_x0_y)$, $p(1_s0_x0_y)$, and $p(1_s1_x0_y)$; the lower row displays the total leakage rate.}\label{fig:leak IBM}
\end{figure}

\begin{figure}
\centering
\includegraphics[width=\textwidth]{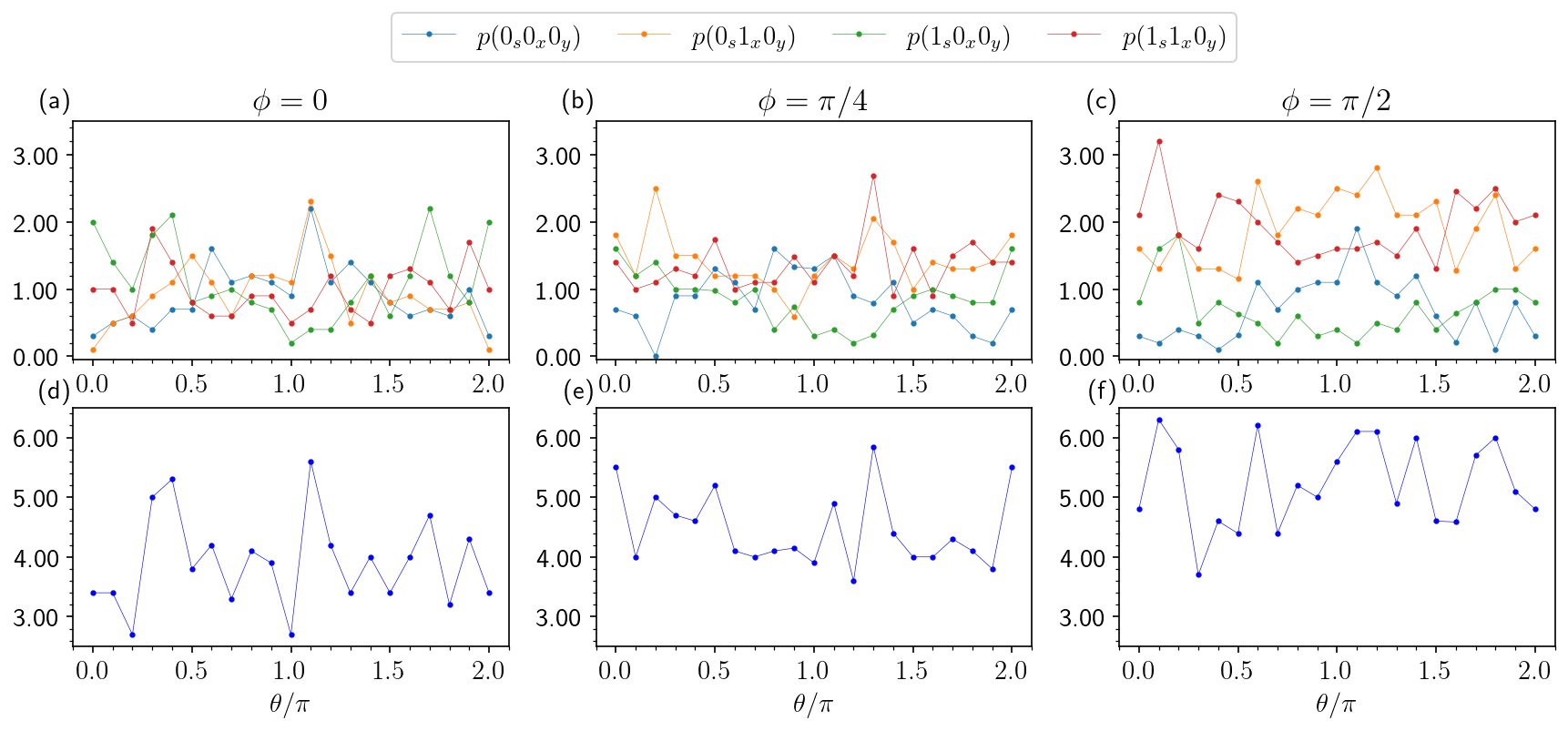}
\caption{Leakage probabilities (\%) of the experiment presented in \figref{fig:ionQ_interference} on \texttt{ionq\_harmony} without the random choice. The upper row display individual probabilities for $p(0_s0_x0_y)$, $p(0_s1_x0_y)$, $p(1_s0_x0_y)$, and $p(1_s1_x0_y)$; the lower row display the total leakage rate.}\label{fig:leak IonQ}
\end{figure}

As noted in the paragraph after \eqref{p xy}, the outcomes $0_x0_y$ and $1_x0_y$ are not expected to occur; however, they still appear as leakage errors.
We characterize the leakage observed on the IBM Quantum and IonQ platforms in \figref{fig:leak IBM} and \figref{fig:leak IonQ}, respectively, for the experiments without random choice, as previously shown in \figref{fig:ibm_interference} and \figref{fig:ionQ_interference}. These figures display the probabilities $p(0_s0_x0_y)$, $p(0_s1_x0_y)$, $p(1_s0_x0_y)$, and $p(1_s1_x0_y)$, as well as the total leakage rate, defined as their sum. The leakage probabilities for other experiments with random choice remain of the same order of magnitude.

\subsection{Transpiled low-level circuits on \texttt{ibm\_kyiv}}
Before a quantum circuit is executed on \texttt{ibm\_kyiv}, it is transpiled into a corresponding low-level circuit in terms of primitive gates: $R_z$ rotation gates, $\sqrt{X}$ gates, $X$ gates, echoed cross-resonance (ECR) gates, and delay gates.

\begin{figure}
\centering
    \includegraphics[width=\textwidth]{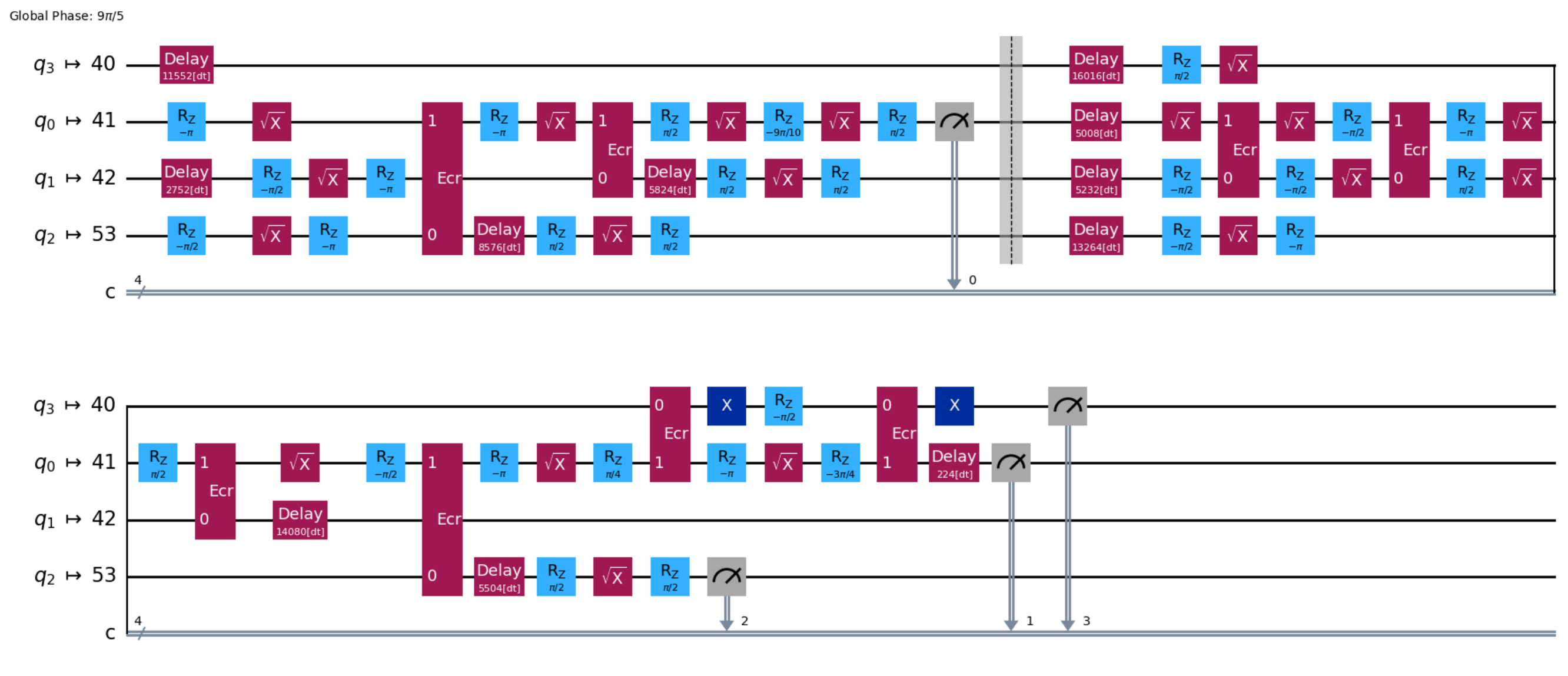}
    \includegraphics[width=\textwidth]{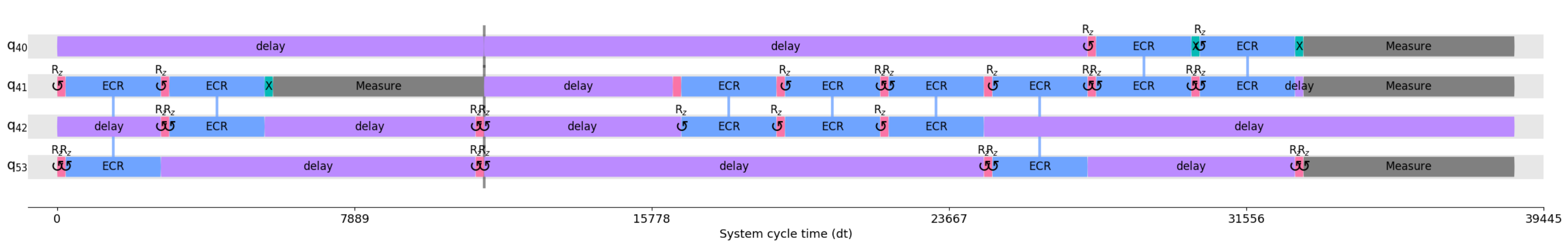}
    \caption{The transpiled circuit on \texttt{ibm\_kyiv} and its detailed execution schedule for \figref{fig:transpiled}, specifically for $\theta = \pi/10$, $\phi = \pi/2$, and $t_{\text{delay}} = 5{,}000\,dt$.}
    \label{fig:low level 2options delay5000}

\end{figure}

The transpiled circuit for \figref{fig:transpiled}, along with its detailed execution schedule, is shown in \figref{fig:low level 2options delay5000}, specifically for $\theta = \pi/10$, $\phi = \pi/2$, and $t_{\text{delay}} = 5{,}000\,dt$, with the initial qubit mapping $(s,x,y,a) \mapsto (41,42,53,40)$.
In the schedule chart, $\sqrt{X}$ gates are represented by pink ribbons; $X$ gates are represented by green ribbons marked with an ``X''; ECR gates are shown as linked pairs of blue ribbons; and $R_z$ rotation gates, which are implemented as ``virtual'' gates in hardware with zero duration \cite{RZGate}, are indicated by circular arrows.

\begin{figure}
\centering
    \includegraphics[width=\textwidth]{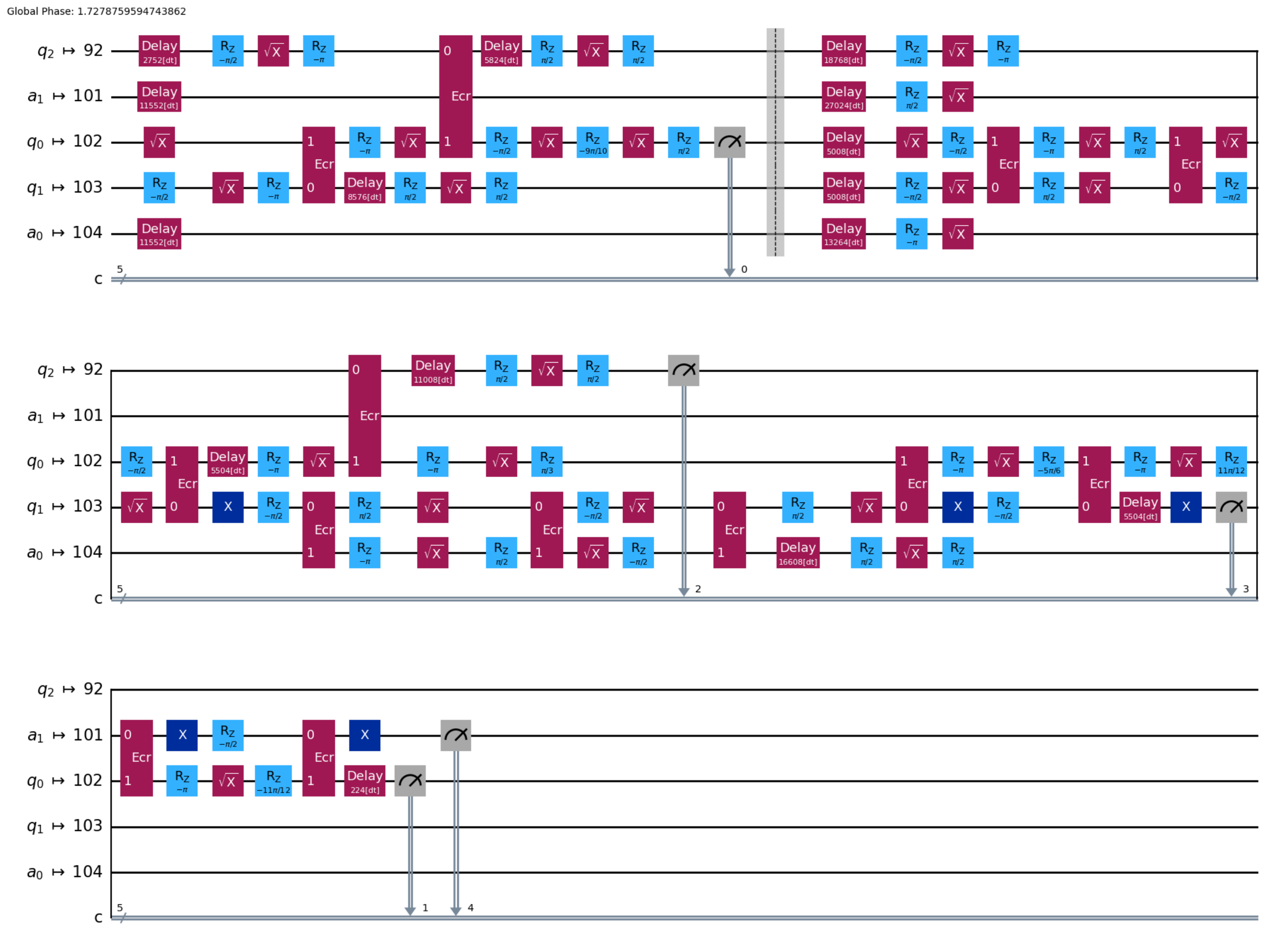}
    \includegraphics[width=\textwidth]{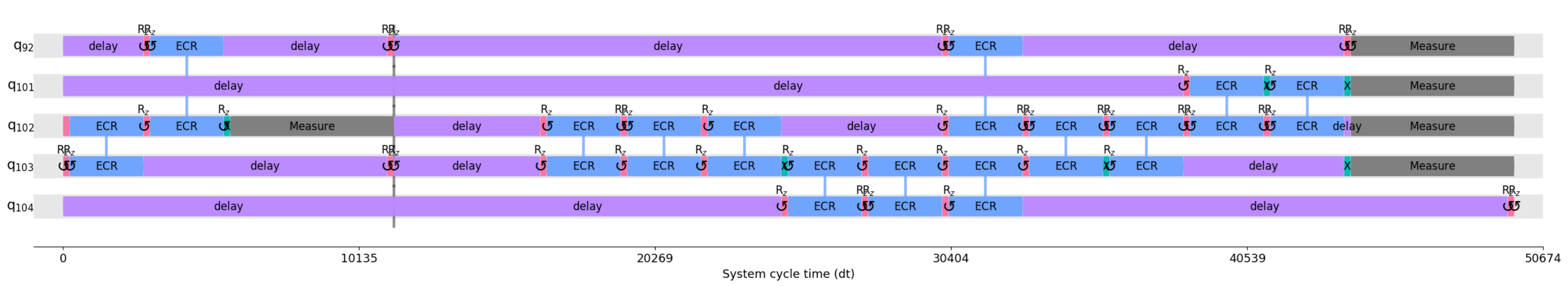}
    \caption{The transpiled circuit on \texttt{ibm\_kyiv} and its detailed execution schedule for the circuit with four-option random choice, specifically for $\theta = \pi/10$, $\phi_1 = \pi/6$, $\phi_2=\pi/3$, and $t_{\text{delay}} = 5{,}000\,dt$.}
    \label{fig:low level 4options delay5000}

\end{figure}

The transpiled circuit for the circuit with four-option random choice, as discussed in \appref{app:IBM 4 options}, along with its detailed execution schedule, is shown in \figref{fig:low level 4options delay5000}, specifically for $\theta = \pi/10$, $\phi_1 = \pi/6$, $\phi_2=\pi/3$, and $t_{\text{delay}} = 5{,}000\,dt$, with the initial qubit mapping $(s,x,y,a_1,a_2) \mapsto (102,103,92,104,101)$.

\subsection{Statistical and sampling uncertainties}

If one performs a sequence of $n$ independent \emph{Bernoulli trials} (also called \emph{binomial trials}), each with an identical probability $p$ of yielding ``success'' and a probability $1-p$ of yielding ``failure'', then the total number of successes is described by the binomial distribution with a mean given by $Np$ and a standard deviation given by $\sqrt{Np(1-p)}$. Equivalently, the random variable representing the total number of successes divided by $N$ is described by a rescaled binomial distribution with a mean $p$ and a standard deviation.
\begin{eqnarray}\label{sigma th}
\sigma_\text{th} = \sqrt{\frac{p(1-p)}{N}},
\end{eqnarray}
which quantifies theoretical statistical fluctuations of the total number of successes divided by $N$ from the fiducial value $p$.

Given a quantum circuit, the probability of obtaining a certain desired measurement outcome is theoretically given by a fixed number $p$, while the probability of obtaining any other outcome is $1-p$. To experimentally estimate the value of $p$, we run the circuit for $N$ shots, where $N$ is a large number. If we observe $N_s$ events for the desired outcome and $N_f = N - N_s$ events otherwise, the best estimate for $p$ is the sample mean:
\begin{eqnarray}
\bar{x} = \frac{N_s}{N}.
\end{eqnarray}
If the sampling process were repeated infinitely many times, each with $N$ shots, we would obtain a distribution of sample means, which has its own variance. This variance is best estimated by the standard error of the mean (SEM):
\begin{eqnarray}\label{SEM}
\sigma_{\bar{x}} = \sqrt{\frac{\sum_{i=1}^N(x_i-\bar{x})^2}{N(N-1)}} = \sqrt{\frac{N_s N_f}{N^2 (N-1)}},
\end{eqnarray}
where $x_i = 1$ if the $i$-th shot yields the desired outcome and $x_i = 0$ otherwise.
The SEM $\sigma_{\bar{x}}$ quantifies the sampling uncertainty regarding how closely the sample mean $\bar{x}$ obtained from sampled shots approximates the true value of $p$. As the number $N$ of shots increases, $\sigma_{\bar{x}}$ decreases, indicating that the estimate $\bar{x}$ becomes more reliable. However, in practice, experimental results are influenced by various errors and noise, leading to greater uncertainty in $\bar{x}$ than what $\sigma_{\bar{x}}$ alone suggests.

For a more detailed discussion, see \cite{altman2005standard}.


\bibliography{bibref}
\end{document}